%% file: paper_arxiv.tex
\documentclass[3p,sort&compress]{elsarticle}
\journal{Nuclear Physics B}
\usepackage{graphicx}
\usepackage{dsfont}
\usepackage{color}
\usepackage{bbm}
\usepackage{pgf}

\newcommand{\correxp}{\kappa}
\graphicspath{{./figs/}}
\makeatletter
\def\input@path{{./figs/}}
\makeatother

\input{format}

\numberwithin{equation}{section}

\begin{document}

\title{Bridges in the random-cluster model}

\author[cov,mon]{Eren Metin El\c{c}i}
\ead{eren.metin.elci@gmail.com}

\author[cov]{Martin Weigel}
\ead{martin.weigel@coventry.ac.uk}

\author[cov]{Nikolaos G.\ Fytas}
\ead{nikolaos.fytas@coventry.ac.uk}

\address[cov]{Applied Mathematics Research Centre, Coventry
University, Coventry, CV1 5FB, England}
\address[mon]{School of Mathematical Sciences, Monash University, Clayton, Victoria 3800, Australia}
\begin{abstract}
  The random-cluster model, a correlated bond percolation model, unifies a range of
  important models of statistical mechanics in one description, including independent
  bond percolation, the Potts model and uniform spanning trees. By introducing a
  classification of edges based on their relevance to the connectivity we study the
  stability of clusters in this model. We prove several exact relations for general
  graphs that allow us to derive unambiguously the finite-size scaling behavior of
  the density of bridges and non-bridges. For percolation, we are also able to
  characterize the point for which clusters become maximally fragile and show that it
  is connected to the concept of the bridge load. Combining our exact treatment with
  further results from conformal field theory, we uncover a surprising behavior of
  the (normalized) {\em variance\/} of the number of (non-)bridges, showing that it
  diverges in two dimensions below the value
  $4\cos^2{(\pi/\sqrt{3})}=0.2315891\cdots$ of the cluster coupling $q$. Finally, we
  show that a partial or complete pruning of bridges from clusters enables estimates
  of the backbone fractal dimension that are much less encumbered by finite-size
  corrections than more conventional approaches.
\end{abstract}

\begin{keyword}
  Potts model \sep critical phenomena \sep percolation theory

  \PACS 64.60.F- \sep 05.70.Ln \sep 05.50.+q
\end{keyword}

\maketitle

%%%%%%%%%%%%%%%%%%%%%%%%%%%%%%%%%%%%%%%%%%%%%%%%%%%%%%%%%%%%%%%%%%%%%%%%%%%%%%%
%%%%%%%%%%%%%%%%%%%%%%%%%%%%%%%%%%%%%%%%%%%%%%%%%%%%%%%%%%%%%%%%%%%%%%%%%%%%%%%
\section{Introduction}

Percolation is probably the most widely discussed and arguably the simplest model of
critical phenomena. Due to a combination of conceptual simplicity and wide
applicability which is a signature of a problem of very general interest, its many
incarnations including bond and site percolation on a lattice as well as continuum
and non-equilibrium, directed variants, have been the subject of thousands of studies
\cite{stauffer:book}. In statistical physics and mathematics alike, the quest to
understand aspects of the percolation problem has led to developments of powerful and
beautiful new techniques \cite{grimmett:perc,bollobas:06}. While the problem is well
defined and interesting for any graph or lattice, both finite and infinite, the
understanding of non-trivial cases is most advanced in two dimensions. There, the
scaling limit of critical percolation can be related to the Coulomb gas
\cite{nienhuis:domb} and conformal field theory \cite{cardy:domb}, leading to exact
results for most critical exponents and certain correlation functions. More recently,
a rigorous approach to conformal field theory was pioneered by Schramm who used a
mapping introduced by L\"owner to construct a way of generating conformally invariant
fractal random curves, the Stochastic (or Schramm) L\"owner Evolutions (SLEs)
\cite{schramm:00}, for which a range of properties, including fractal dimensions, can
be calculated exactly. In this context, Smirnov and co-workers used the concept of
discrete analyticity to establish rigorously that the scaling limits of critical
percolation \cite{smirnov:01,smirnov:01a} and the Ising model \cite{duminil:12} on
the triangular lattice are indeed conformally invariant, and cluster boundaries in
these models converge to certain classes of SLE traces.

The random-cluster (RC) model was suggested by Fortuin and Kasteleyn as a natural
extension of the (bond) percolation problem, noting that there was a class of models
fulfilling the series and parallel laws of electrical circuits that also included the
Ising model \cite{fortuin:72a}. Given a graph $G=(V,E)$, it assigns to a spanning
subgraph $(V,A\subseteq E)$ a probability mass (in the following also referred to as
RC measure) \cite{grimmett:book}
\begin{equation}
  \probrcm[A] = \frac{p^{\vert A\vert } (1-p)^{\vert E \vert - \vert A \vert }q^{K(A)}}{Z_{\rm RC}(p,q,G)}, \quad A \in \Omega_G,
    \label{eq:rcmprob}
\end{equation}
where $K(A)$ is the number of components and $\vert A\vert$ the number of edges in
$A$. The quantity $Z_\mathrm{RC}(p,q,G)$ is the partition function of the RC model,
corresponding to the sum of unnormalized weights, and $\Omega_G$ constitutes the set of
all spanning subgraphs or configurations, i.e., $\Omega_G = \{ A : A \subseteq E
\}$. Edges that are in $A$ are called \textbf{open} and those in $E \setminus A$
\textbf{closed}. The presence of the cluster-weight factor $q^{K(A)}$ distinguishes
(\ref{eq:rcmprob}) from the percolation problem and prevents $\probrcm[A]$ from being
a Bernoulli product measure; only for $q\to 1$, where the model reduces to the
percolation problem and hence edges become independent, this property is
restored. Although the cluster weight $q$ can be any non-negative real number,
integer values of $q$ are particular in that for them the partition function is very
closely related to that of the $q$-state Potts model \cite{wu:82a} (see
Eq.~(\ref{eq:identity_of_partition_functions}) below). For lattice graphs in at least
two dimensions, the model undergoes a percolation phase transition at a critical
value $p_c(q)$ of the bond probability where, for sufficiently large $q$ the
transition becomes discontinuous \cite{grimmett:book}. On the square lattice,
self-duality allows to deduce the exact transition point $p_\mathrm{c}(q) =
p_\mathrm{sd}(q) = \sqrt{q}/(1+\sqrt{q})$ \cite{beffara:12}, and the location of the
tricritical point is known to be $q_\mathrm{c} = 4$, beyond which the transition
becomes of first order \cite{baxter:book}.

The crucial importance of the RC description for the understanding of critical
phenomena is through its expression in purely geometrical terms. Hence understanding
the geometric structure of the (correlated) percolation problem (\ref{eq:rcmprob})
provides a geometric route to the understanding of the thermal phase transition of
the Potts model. Correspondingly, significant effort has been devoted in particular
to investigations of the structure of the incipient percolating cluster. While
initially it was assumed that it was a network of {\em nodes\/} connected by
essentially one-dimensional {\em links\/}, results regarding the conductivity of the
critical cluster implied that, instead, the structure is better described by the more
elaborate `links--nodes--blobs' picture \cite{stanley:77}. If one fixes two distant
points $A$ and $B$ on the cluster, those bonds that have independent,
non-intersecting paths to both $A$ and $B$ form the {\bf backbone\/} of the
cluster. It is essentially the part that would carry current in case a voltage was
applied between $A$ and $B$. The remaining cluster mass is in {\bf dangling\/}
ends. The backbone itself consists of singly-connected or {\bf red bonds\/} that
destroy percolation or, equivalently, destroy conductivity between $A$ and $B$ if cut
(the {\bf links} which meet in the {\bf nodes}) as well as bonds in cycles that are multiply
connected (the {\bf blobs}).  These subsets of bonds have fractal scaling with associated
exponents $d_\mathrm{F}$ for the cluster mass itself, $d_\mathrm{BB}$ for the
backbone, and $d_\mathrm{RB}$ for the red bonds. Obviously one has $d_\mathrm{RB} \le
d_\mathrm{BB} \le d_\mathrm{F}$, and it turns out that strict inequality holds in the
generic case and hence asymptotically almost all of the bonds of the incipient
percolating cluster are in dangling ends, and almost all of the mass of the backbone
is in the blobs. In two dimensions, exact expressions for $d_\mathrm{F}$ and
$d_\mathrm{RB}$ are available from Coulomb gas arguments \cite{coniglio:89}, but
$d_\mathrm{BB}$ is only known numerically \cite{deng:04,xu:14}.

As we discuss here, the distinction of singly-connected and multiply-connected bonds
first suggested for understanding the backbone structure \cite{stanley:77} is a more
generally useful classification of bonds in a cluster configuration. Removing singly
connected bonds, which we call {\bf bridges\/},\footnote{Note that in some of the
  previous literature on percolation, the term {\bf bridge\/} is used to denote the
  (candidate) red bonds \cite{coniglio:89,araujo:11a}. While these are actually {\em
    backbone bridges\/}, we use the term in a more general sense here to denote
  singly-connected bonds anywhere in the cluster configuration.} generates an
additional connected component, while this is not the case for multiply connected
bonds or {\bf non-bridges\/}. Indeed, this separation allows us to control the effect
of a local edge manipulation (open $\leftrightarrow$ closed) on the weight of the
configuration as in Eq.~(\ref{eq:rcmprob}). We will exploit this below to derive
expressions for the expected number of bridges and related classes of bonds as well
as the scale of fluctuations in the former. A further partitioning of bridges was
recently suggested for percolation in Ref.~\cite{xu:14}: {\bf branches\/} are bridges
for which at least one of the connected clusters is a tree, whereas the remaining
bridges are dubbed {\bf junctions\/}. This distinction is related to but not
congruent with the links--nodes--blobs picture: most of the red bonds will be
junctions, but so are some bonds in dangling ends as well. Cutting the branches, on
the other hand, will remove some part of the dangling ends (the treelike ones), and
branches are not typically found in the backbone\footnote{This is a plausible
  assumption as the defining property of a branch requires that at least one end of the
  considered edge is a tree, which in the presence of blobs on the backbone is
  unlikely in the sense of an asymptotically vanishing density.}. As a result of this
incongruence, the number of junctions on the percolating cluster grows as
$L^{d_\mathrm{F}}$ and not proportional to $L^{d_\mathrm{RB}}$, where $L$ is the
linear size of the system. In contrast to the class of red bonds, therefore, both
branches and junctions are extensive subsets of the bridges. For uncorrelated
percolation Xu {\em et al.\/} \cite{xu:14} find numerically, however, that removing
all bridges, i.e., both branches and junctions, leads to tightly connected
remainders, corresponding to the blobs, whose size grows as $L^{d_\mathrm{BB}}$. As
we will show below in Sec.~\ref{sec:bridge_free}, it suffices to remove the junctions
only to see the same behavior more generally for the RC model.

As, by definition, the removal of a bridge leads to the generation
of a new component and hence the  breakup of an existing cluster,
understanding the properties of bridge bonds is crucial also to
the understanding of fragmentation phenomena in the framework of a
lattice model. This connection was investigated in our recent
letter \cite{elci:15}, where we studied the fragmentation rate and
kernel, and related the associated scaling exponents to the more
standard critical exponents and fractal dimensions. In the present
paper, however, we follow a different line of study. In
particular, we derive a linear relation between the expected
density of bridges and the density of bonds; we provide the
asymptotic densities of bridges in the thermodynamic limit for the
square lattice and study the scaling corrections in finite systems
in general. We employ Monte Carlo simulations based on the
Swendsen-Wang--Chayes-Machta algorithm
\cite{swendsen-wang:87a,chayes:98a} and a recent new
implementation \cite{elci:13} of Sweeny's algorithm
\cite{sweeny:83} to confirm these results for the square lattice
and study the behavior in three dimensions. Building on these
results for the densities, i.e., the first moment of the bridge
distribution, we move on to studying the fluctuations,
corresponding to the second moment. We reveal an unexpected
finite-size behavior leading to a non-specific-heat variance singularity for $q\leq
4\cos{(\pi/\sqrt{3})}^2$, a direct consequence of an intriguing
interplay between bridges and non-bridges.

The rest of this paper is organized as follows. In
Section~\ref{sec:bridge_density} we introduce the edge
classification, derive the bridge-edge identity in general,
consider some of its consequences and present a numerical study
for the special case of the square lattice, together with a
detailed analysis of finite-size corrections. In
Section~\ref{sec:candidatebridges} we use symmetries of the RC
model in order to derive connections to other quantities of
interest in the percolation literature, such as the short range
connectivity. In Section~\ref{sec:bridge_maximum} we introduce the
concept of bridge load which allows us to analytically
characterize the point of maximal bridge density for independent
bond percolation. In Section~\ref{sec:bridge_correlations} we
derive a second-order bridge-edge identity that allows us to study
the variance of the number of bridges. In
Section~\ref{sec:bridge_free} we compare various new and old
methods to extract the backbone fractal dimension, with an
emphasis on the involved finite-size corrections.
Section~\ref{sec:conclusion} contains our conclusions.

\section{Bridge density\label{sec:bridge_density}}

\subsection{Edwards-Sokal coupling}

Although in this paper we focus on the RC model itself, its
relation to the Potts model is of relevance for the physical
interpretation of the (geometric) percolation transition in terms
of a thermal phase transition. In particular, we make reference to
how certain observables of the Potts model translate into
quantities in the RC language. The Potts model describes
interacting spins, and assigns one of the $q$ spin values in
$\{1,\ldots,q\}$ to each vertex, attributing a configurational
probability or Boltzmann weight
\begin{equation}
    \mathbb{P}_{\beta,q,G}[\boldsymbol{\sigma}] \equiv \frac{1}{Z_\mathrm{P}(\beta,q,G)}
    \sum_{\boldsymbol{\sigma} \in \mathcal{Q}}
    \exp{\left(\beta \sum_{(x,y) \in E} \indicator_{
          \{\sigma_x = \sigma_y\}} \right)} =
    \frac{1}{Z_\mathrm{P}(\beta,q,G)} \prod_{(x,y) \in E} e^{\beta}[1-p+p\indicator_{
          \{\sigma_x = \sigma_y\}}]
    ,
    \label{eq:potts}
\end{equation}
to a spin configuration $\boldsymbol{\sigma} = \{\sigma_1,\ldots,\sigma_{\vert V \vert}\} \in
\mathcal{Q} = \{1,\cdots,q\}^{\vert V \vert}$. Here, we write $\indicator_{\{\sigma_x =
  \sigma_y\}}$ for the indicator function, which equals $1$ if $\sigma_x = \sigma_y$
and $0$ otherwise. Analogous to $Z_\mathrm{RC}(p,q,G)$, the Potts partition function
$Z_\mathrm{P}(\beta,q,G)$ corresponds to the sum of the unnormalized
weights. Importantly, in the second equality, we need to identify $p =
1-e^{-\beta}$. The relation between Eqs.~(\ref{eq:rcmprob}) and (\ref{eq:potts}) is
established by augmenting phase space to include both, spin variables
$\boldsymbol{\sigma}$ and subgraph variables $A$, leading to the joint
probability
\begin{equation}
  \mathbb{P}_{p,q,G}[\boldsymbol{\sigma}, A] \equiv \frac{1}{Z_\mathrm{joint}(p,q,G)}
   \prod_{(x,y) \in E} e^{\beta} [(1-p)\indicator_{\{e\notin A\}}+p\indicator_{
          \{\sigma_x = \sigma_y\}} \indicator_{\{e \in A\}}],
      \label{eq:EScoupling}
\end{equation}
that is also known as Edwards-Sokal coupling \cite{edwards:88a}.
It has the crucial property that the marginal distribution on the
spins alone reduces to the Potts weight (\ref{eq:potts}), while
the marginal on the subgraphs gives the RC weight
(\ref{eq:rcmprob}). Similarly, one can derive the conditional
probabilities for the spins given the bonds or vice versa
\cite{grimmett:book}. As was already shown by Fortuin and
Kasteleyn \cite{fortuin:72a}, the corresponding partition
functions coincide up to a (trivial) multiplicative factor if we relate the
parameters $\beta$ and $p$ via $p = 1-e^{-\beta}$,
\begin{equation}
    Z_{\rm RC}{\left(1-e^{-\beta},q,G\right)} = e^{-\beta \vert E \vert} Z_{\rm P}(\beta,q,G)~\textnormal{for}~q \in \{2,3,\cdots\}.
    \label{eq:identity_of_partition_functions}
\end{equation}
This coupling is the basis for the Swendsen-Wang cluster algorithm
\cite{swendsen-wang:87a} and its generalization to non-integer $q$ due to Chayes and
Machta \cite{chayes:98a}. The above coupling and the identity
(\ref{eq:identity_of_partition_functions}) allow in particular to relate expectation
values in the two ensembles. For instance, the internal energy in the Potts model is
up to constants identical to the expected number of edges in the RC language,
\begin{equation}
  \label{eq:energy_vs_bonds}
  u_{\beta,q,G} \equiv -\frac{1}{\vert V \vert} \frac{\partial \ln Z_\mathrm{P}(\beta,q,G)}{\partial
    \beta} = -\frac{1}{p}\frac{\vert E \vert}{\vert V \vert} \avrcm[\mathcal{N}],~\textnormal{for}~p = 1-e^{-\beta},
\end{equation}
where ${\cal N}(A) = |A|/\vert E \vert$ is the proportion of open edges and $\avrcm[\cdot]$
denotes the expectation with respect to the RC measure (\ref{eq:rcmprob}). Similarly,
higher moments of the energy distribution such as the specific heat as well as
magnetic observables of the Potts model can be expressed in terms of expectation
values in the RC language \cite{grimmett:book}. Due to the conditional measures,
correlation functions can also be related. Consider, in particular, the
event $x \stackrel{A}{\leftrightarrow }y$, that is the existence  of
a path in $A$ connecting $x$ and
$y$. From the conditional measures derived from Eq.~(\ref{eq:EScoupling}) it is easy
to see that
\begin{equation}
  \label{eq:connectivity}
  \avrcm[1-\indicator_{\{\sigma_x = \sigma_y\}} | A] =
  \frac{q-1}{q}\left[1- \indicator_{\left\{x \stackrel{A}{\leftrightarrow }y\right\}}
  \right].
\end{equation}
That is two vertices can only have unequal spin when they are not connected in the
bond configuration and were assigned two different (random) colors. Summing over the
edges shows that the nearest-neighbor connectivity defined as the sum
\begin{equation}
  \label{eq:connectivity2}
  {\cal E}(A) \equiv \frac{1}{\vert E \vert} \sum_{(x,y)\in E}
  \indicator_{\left\{x \stackrel{A}{\leftrightarrow }y\right\}}
\end{equation}
is essentially the energy \cite{salas:97,grimmett:book,hu:14}
\begin{equation*}
  u_{\beta,q,G} = \frac{\vert E\vert}{\vert V \vert} - \frac{q-1}{q} \left(\frac{\vert E \vert}{\vert V \vert}  -
    \avrcm[\mathcal{E}]\right)~\textnormal{for}~p=1-e^{-\beta}.
\end{equation*}
Clearly for hypercubic lattices $\vert E \vert/\vert V \vert$ is
bounded (more precisely equals $d$, where $d$ is the
dimensionality).

\subsection{Edge classification}

Intuitively, \eqref{eq:rcmprob} indicates that for $q> 1$ $(q<1)$ the model favors
configurations with a larger (smaller) number of components, compared to the case
$q=1$. In this paper we investigate how the cluster structure is influenced by the
cluster weight $q$, in particular with respect to the fragility of clusters. The
relevance of individual edges for the connectivity of $A\subseteq E$ depends on
whether they are \textbf{pivotal} or \textbf{non-pivotal}. According to our definition,
pivotal edges change the number of components upon removal/insertion, i.e., for a
pivotal edge $e$ we have $K(A^e) = K(A_e)-1$, where $A^e$ ($A_e$) is the
configuration obtained from $A$ by opening (closing) $e$ (note that necessarily one
of $A^e$ and $A_e$ is equal to $A$).  Hence for a pivotal edge there is no
alternative path connecting the vertices incident to it. As discussed above, for the
purposes of this paper open pivotal edges are denoted as \textbf{bridges}, while we
call pivotal closed edges \textbf{candidate-bridges}. Likewise, we denote open (closed)
non-pivotal edges as \textbf{non-bridges} (\textbf{candidate-non-bridges}).  Let $B$,
$C$, $\overline{B}$, $\overline{C}$ denote the set of bridges, non-bridges,
candidate-bridges and candidate-non-bridges, respectively. We stress that these sets
depend explicitly on the configuration $A$. We denote the corresponding densities as
${\cal B} = |B|/\vert E \vert$, ${\cal C} = |C|/\vert E \vert$, $\overline{\cal B} = |\overline{B}|/\vert E \vert$, and
$\overline{{\cal C}} = |\overline{C}|/\vert E \vert$.

\subsection{Derivation of the bridge-edge formula}

In this sub-section we show how, for \emph{arbitrary} (finite)
graphs, the expected densities of all the above mentioned
edge-types $B$, $C$, $\overline{B}$, and $\overline{C}$ can be
related to the expectation of the edge density. We establish these
identities by utilizing a differential equation known as
Russo-Margulis formula \cite{grimmett:graphs,steif:11}, which
applies to expectations with respect to product measures such as
the probability density of the bond percolation model, corresponding
to the $q=1$ RC model. The formula intuitively quantifies the
``response'' to an infinitesimal change in the bond density. Using
this approach to quantify the response in the number of connected
components $K$ allows us to study the bridge density. We then
proceed and apply the idea to the RC model with generic $q>0$, by
exploiting the fact that the partition function of the model can
be interpreted as a particular expectation in the percolation
model.

First consider the percolation model, i.e., the measure (\ref{eq:rcmprob}) with
$q=1$, for which we denote expectations as $\avperc[\cdot]$. In this setting, the
Russo-Margulis formula reads
\begin{equation}
  \frac{\mathrm{d}}{\mathrm{d}p} \avperc[X] = \sum_{e \in E}\avperc[\delta_e X],
  \label{eq:russo}
\end{equation}
where $X(A): \Omega \rightarrow \mathbb{R}$ is an arbitrary observable and the
quantity $\delta_e X$ is called the \textbf{influence}  of $e$ on $X$ (or the
discrete $e$-derivative of $X$) and is given by
\begin{equation*}
  (\delta_e X)(A) \equiv X(A^e) - X(A_e).
\end{equation*}
%We emphasize that the Russo-Margulis formula is not restricted to graphs, and is applicable to
%product measures in general.
The graph specific geometric properties of the model under study are ``encoded'' in
the influences.  In order to simplify the subsequent analysis, we show in Appendix
\ref{ap:a} the following bijection identity:
\begin{equation}
  \avperc[\delta_e X] = \frac{1}{p} \sum_{A\subseteq E : \atop e \in A} \probperc[A]
  \left[ X(A) - X(A_e) \right].
  \label{eq:inflbridge}
\end{equation}
We now apply this identity to the cluster-number observable $K$, for which
$K(A)-K(A_e) = -1$ if $e$ is a bridge, $e\in B(A)$, and $0$ otherwise. Application of
the Russo-Margulis formula \eqref{eq:russo} and Eq.~\eqref{eq:inflbridge} to $X=K$
then yields
\begin{equation}
  \frac{\mathrm{d}}{\mathrm{d}p}\avperc[K] = \sum_{e \in E} \avperc[\delta_e K]
  = \frac{1}{p} \sum_{e \in E} \sum_{A \subseteq E: \atop e \in A} \probperc[A] \left[ K(A) - K(A_e) \right]
  = -\frac{1}{p} \sum_{e \in E} \probperc[e\in B]
  = -\frac{\vert E \vert}{p} \avperc[\bridge].
 \label{eq:rsfmlperc} 
  \end{equation}
This differential equation allows us to
determine the cluster numbers $\avperc[K]$ once the $p$-dependence of
$\avperc[\mathcal{B}]$ is known. We remark that this relation was derived before in
\cite{essam:87}, however with a different target application in mind. As already mentioned,
we can extend the above to the RC model and obtain more generally the bridge density
$\avrcm[\mathcal{B}]$ of the RC model itself. To start with, we note that the RC
model partition function is $Z_{\rm RC}(p,q,G) = \avperc[q^K]$. In other words,
the RC model partition function is the expected value of $q^K$ in the bond percolation
model with parameter $p$. Hence we can
apply the Russo-Margulis formalism to $q^K$. Using the same
reasoning as that for $\avperc[K]$ above, we show in Appendix~\ref{ap:b} that
\begin{equation}
    \frac{\partial}{\partial p} \log{Z_{\rm RC}(p,q,G)} = \vert E \vert\frac{1-q}{p} \avrcm[\mathcal{B}].
    \label{eq:der_freeenergy}
\end{equation}
On the other hand, direct differentiation of the partition function $Z_\mathrm{RC}(p,q,G)$ yields
\begin{equation}
    \frac{\partial}{\partial p} \log{Z_{\rm RC}(p,q,G)} = \frac{\vert E\vert}{1-p} \left(\frac{1}{p} \avrcm[\mathcal{N}]-1\right).
    \label{eq:der_freeenergy_2}
\end{equation}
Comparing both expressions \eqref{eq:der_freeenergy} and \eqref{eq:der_freeenergy_2},
we derive the following relationship for the RC model (referred to as
\textbf{bridge-edge formula}),
\begin{equation}
    \avrcm[\bridge] = \frac{\avrcm[\edge] - p}{(1-p)(1-q)},
    \label{eq:avbridges}
\end{equation}
which is of central importance for the following analysis (see \cite{caselle:01} for
some related identities developed along different lines). We remark again that this
holds for any graph and for any $p\in (0,1)$ and $q\in (0,\infty) \setminus 1$ (the
cases $q\rightarrow 0$, $1$ are discussed below). Due to the obvious identity
$\avrcm[\edge] = \avrcm[\bridge] + \avrcm[\nbridge]$, one immediately arrives at an
analogous equation for the density of non-bridges,
\begin{equation}
    \avrcm[\nbridge] = \frac{(pq-p-q)\avrcm[\edge] - p}{(1-p)(1-q)}.
    \label{eq:avnonbridges}
\end{equation}
We remark that in principle one can use the Russo-Margulis
formalism to study the (partial) $p$-derivative of expectations of
other observables in the RC model:
$$
\frac{\partial}{\partial p} \avrcm[X] =
\frac{\partial}{\partial p} \left( \frac{\avperc[q^K
X]}{\avperc[q^K]} \right) = \frac{\frac{\partial}{\partial p}\avperc[q^K
X]}{\avperc[q^K]} + \frac{q-1}{p} \avrcm[X]
\avrcm[\vert B \vert],
$$
which shows the relevance of bridges in general for the RC model.
Furthermore, let us consider a few direct consequences and
applications of \eqref{eq:avbridges}:
\begin{enumerate}
\item Firstly, notice that \eqref{eq:avbridges} recovers the correct results for the
  limit of uniform spanning trees, where all open edges are bridges.  This tree model
  can be obtained in the RC model in the limit $q,p\rightarrow 0$ for
  $q/p\rightarrow 0$, which for the square lattice includes the critical line $p_{\rm
    sd}(q)$ \cite{grimmett:book}.
%\item Consider the set of all connected spanning subgraphs of a connected graph $G$.  It
%  is not too hard to see that in the limit $q\rightarrow 0$ with $p$ fixed, the
%  RC model reduces to the weighted connected spanning subgraph model \cite{jacobsen:04},
%  where each connected spanning subgraph $A$ has weight $\left(p/(1-p)\right)^{\vert
%    A \vert}$. For this case Eq.~\eqref{eq:avbridges} reduces to
%  \[
%  \mathbb{E}_{p,0}[\bridge] = \frac{\mathbb{E}_{p,0}[\edge] - p}{1-p}.
%  \]
%  It is clear that by setting $p=1/2$ each supported configuration carries
%  the same probability mass, that is $p=1/2$ corresponds to the uniform weight
%  over all connected spanning subgraphs. In this case one obtains $\mathbb{E}_{1/2,0}[\bridge] =
%  2\mathbb{E}_{1/2,0}[\edge] - 1$. This relationship appears to be a novel nontrivial result.
\item It is possible to establish an upper bound for $\avrcm[\bridge]$ for $q\geq 1$
  and any $p$ on any graph. Using general comparison inequalities for the
  RC measure \cite{grimmett:95}, it is possible to show that
 \begin{equation*}
    \avrcm[\edge] \geq \mathbb{E}_{\tilde{p}(p,q),1,G}[\edge]
     = \frac{p}{(1-p)q + p}  \equiv  \tilde{p}(p,q),
  \end{equation*}
  where $ \tilde{p}(p,q) \leq p$ for $q\geq 1$.
  Together with \eqref{eq:avbridges} this yields the desired upper bound
  for $\avrcm[\bridge]$
  \begin{equation*}
    \avrcm[\bridge] = \frac{p - \avrcm[\edge]}{(1-p)(q-1)}
    \leq \frac{p - \tilde{p}(p,q)}{(1-p)(q-1)}
    = \tilde{p}(p,q).
  \end{equation*}
  As is shown below for the Ising case $q=2$ with an exact solution this bound is
  tight in the limit $p\to 0$, but clearly not so for $p \gtrsim p_\mathrm{sd}$.
\item Another interesting consequence follows when one recasts (\ref{eq:avbridges})
  to express the edge density in terms of the bridge density
  \begin{equation}
    \avrcm[\edge] =p - (q-1)(1-p)\avrcm[\bridge].
    \label{eq:vanamp}
  \end{equation}
  This emphasizes the importance of bridges for $q\neq 1$ as the source of
  finite-size corrections in $\avrcm[\edge]$, a manifestation of the
  introduction of correlations (deviation from the product measure).  Moreover it
  also explains, in the setting of the RC model, how thermal fluctuations in the number of edges vanish in
  the percolation limit $q\rightarrow 1$, namely by a vanishing factor $(q-1)$. This
  has been a plausible assumption in the literature so far, but to our knowledge,
  relation \eqref{eq:vanamp} is the first exact statement, see for instance
  \cite{hu:14, hu:99}.
\end{enumerate}

\subsection{The square lattice}

For the infinite square lattice $\mathbb{Z}^2$, the self-dual point $p_{\rm sd}(q) =
\sqrt{q}/(1+\sqrt{q})$ is critical\footnote{Note however that albeit being
  numerically and physically well supported, this has only recently been rigorously
  confirmed for the case $q\geq 1$ \cite{beffara:12}.}. For the Potts model at
criticality, a mapping to an exactly-solved six-vertex model gives extensive
analytical results \cite{wu:82a,baxter:book}; it is assumed that by analytic
continuation these carry over to general values of $q$. In particular, the critical
internal energy density in this case is $u_{\beta_c(q),q,\mathbb{Z}^2} = 1+1/\sqrt{q}$ and
hence from Eq.~(\ref{eq:energy_vs_bonds}) one deduces that $\avrcmcsq[\edge] = 1/2$,
where we defined in general $\avrcmc[\cdot] \equiv \mathbb{E}_{p_{c}(q),q}[\cdot]$ for
any graph $G$ (analogous for probabilities), where of course the critical line is given
by $p_c(q)=p_{\rm sd}(q)$ for $\mathbb{Z}^2$. 
It follows from (\ref{eq:avbridges}) and (\ref{eq:avnonbridges})
that one has
\begin{equation}
    \avrcmcsq[\bridge] = \frac{1}{2(1+\sqrt{q})} , \quad
    \avrcmcsq[\nbridge]  = \frac{\sqrt{q}}{2\left(1+\sqrt{q}\right)} \label{eq:asymp}.
\end{equation}
We note that Eqs.~\eqref{eq:asymp} can be written in terms of $p_{\rm sd}(q)$ and
$\avrcmcsq[\edge]$, as
\begin{equation}
\avrcmcsq[\bridge] = \big[1-p_{\rm sd}(q)\big] \avrcmcsq[\edge] , \quad
\avrcmcsq[\nbridge] = p_{\rm sd}(q) \avrcmcsq[\edge].
\end{equation}
The above expressions allow us interpret $p_{\rm sd}(q)$ as the expected fraction of
non-bridges among all open edges. Clearly, the remaining fraction $1-p_{\rm sd}(q)$
of open edges have to be bridges. We note that because of $e\in B \Rightarrow e\in A$
we have $\probrcm[e\in B \vert e\in A] = \probrcm[e \in B]/ \probrcm[e \in A]$ and
hence in particular $\probrcmcsq[e \in B \vert e \in A] = 1-p_{\rm sd}(q)$ and
similarly $\probrcmcsq[e \in C \vert e \in A] = p_{\rm sd}(q)$. Interestingly, an
immediate consequence of \eqref{eq:asymp} is that our RC result reduces in the
percolation limit, $q\rightarrow 1$, to $\avrcmcsq[\bridge]$, $\avrcmcsq[\nbridge]
\rightarrow 1/4$, which recovers a result recently derived in Ref.~\cite{xu:14}.

\begin{figure}[tb]
%scripts: figs/plot_two_rows_bridges.py
\centering
%\the\textwidth
\input{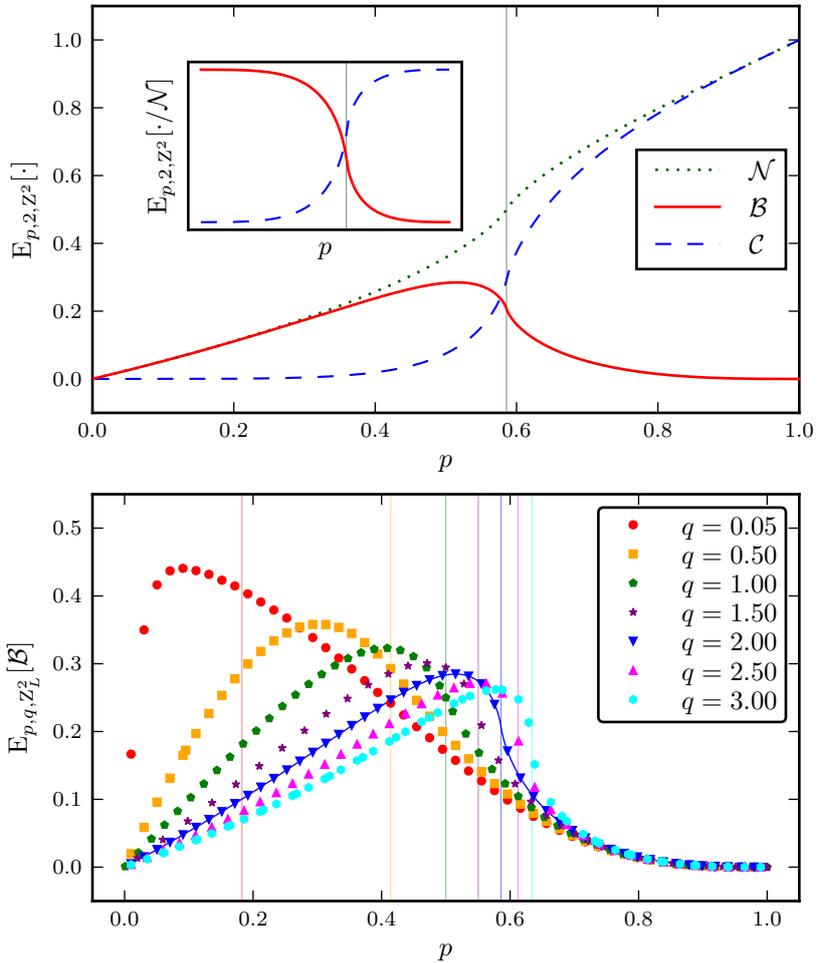}
\caption{Upper panel: Exact densities of edges, bridges and non-bridges for the $q=2$
  RC model on $\mathbb{Z}^2$. The red line shows the bridge density, the blue line
  denotes the non-bridge density and the green curve represents the edge density. The
  inset shows the expected fraction of bridges (red) and non-bridges (blue) among all
  open edges. Lower panel: For comparison we also show numerical estimates
  for the bridge density on $\mathbb{Z}^2_L$ for cluster weights in $[0,4]$ and $L=64$ whenever $q\neq 1$ and
      $L=2048$ for $q=1$. The vertical lines show the
  location of the self-dual (critical) points for the relevant values of $q$.}
    \label{fig:bridge_nonbridge_ising}
\end{figure}

Going beyond criticality, for the Ising model, corresponding to
$q=2$, we can use the available exact solution for the internal
energy and the relation (\ref{eq:energy_vs_bonds}) to find the
exact $p$-dependence of the densities of edges as well as bridges
and non-bridges. This allows us to study, in an exact setting, the
fragility as the bond-density/temperature is varied. As the
corresponding expression for the internal energy is not very
instructive we refrain from reproducing it here, see, e.g.,
Ref.~\cite{baxter:book} for details. Instead, we show in the upper
panel of Figure \ref{fig:bridge_nonbridge_ising} the exact
asymptotic results for $\mathbb{Z}^2$. For comparison, we also
show numerically estimated values of the density of bridges for
other cluster weights (for details regarding the numerical
procedure, see Sec.~\ref{sec:numerical_analysis} below) in the
lower panel of Fig.~\ref{fig:bridge_nonbridge_ising}. As it is
clear from Fig.~\ref{fig:bridge_nonbridge_ising}, the density of
bridges is not maximized at the critical point but for some
$p_{\rm f}(q) < p_{\rm sd}(q)$. We shall give an explanation for
this phenomenon in terms of the relationship between
nearest-neighbor connectivity and bridge density for the
percolation case $q=1$ in Sec.~\ref{sec:percolation_specialities}.
Regarding the results for the Ising model, we note that it is also
possible to produce exact expressions for the internal energy on
finite lattices \cite{ferdinand:69a}, resulting in corresponding
exact results for the bridge densities on $\mathbb{Z}^2_L$, the
$L\times L$ square lattice with periodic boundary conditions.

It is apparent from the lower panel of Fig.~\ref{fig:bridge_nonbridge_ising} that a
characteristic of the critical point is the extremal slope of
$\avrcmsq[\bridge]$. Moreover, the relation \eqref{eq:avbridges} suggests that the
slope becomes singular due to a thermal singularity for $q\geq 2$. Indeed, the
occurrence of a singular slope can be explained by using \eqref{eq:der_freeenergy}
connecting $\avrcm[\bridge]$ to the first $p$-derivative of the $\log{(Z_{p,q})}$,
which allows one to derive
\begin{equation}
\partial_p \avrcm[\bridge] = \frac{ p \partial_p^2 \log{\left[Z_{{\rm RC}}(p,q,G)\right]}
}{\vert E \vert(1-q)} +  \frac{1}{p} \avrcm[\bridge].
\label{eq:deriv_bridge}
\end{equation}
this latter relation together with (\ref{eq:energy_vs_bonds})
yields the asymptotic scaling $\partial_p \mathbb{E}_{p,q,\mathbb{Z}^2_L}[\bridge] \vert_{p=p_{\rm sd}(q)} \approx
\partial^2_{\beta}f\vert_{\beta=\beta_c(q)} \approx \log{L}$ for $q=2$
\cite{baxter:book} and $L^{\alpha/\nu}$ for all other values in $q\in(0,4]$, with
multiplicative logarithmic correction proportional to $\log^{-3/2}{(L)}$ for the
tricritical case $q=4$ \cite{salas:97,salas:97a,nauenberg:80,cardy:80}. Here, $f$ is
the free-energy density. We note that Coulomb gas arguments yield expressions for
$\alpha/\nu$ for $q\in (0,4]$ that predict $\alpha/\nu > 0$ whenever $q>2$
\cite{nienhuis:domb}, implying a finite slope for $q<2$ even in the limit
$L\rightarrow \infty$.  We refer the reader to Section \ref{sec:fss} for more details
about size dependent effects and the universality of the above arguments.

\subsection{The percolation case\label{sec:percolation_specialities}}

Returning to the case of general graphs, we see that the bridge-density formula
(\ref{eq:avbridges}) is singular for $q=1$. To deduce the correct result in the
percolation limit $q \to 1$, we use L'H\^{o}pital's rule to find
\begin{equation}
    \lim_{q\to 1} \avrcm[\bridge] = \frac{1}{1-p}
    \lim_{q\to 1} \frac{p-\avrcm[{\cal N}]}{q-1}
    = -\frac{1}{1-p} \lim_{q\to 1}\frac{\partial}{\partial q}\avrcm[\edge]
    = -\frac{1}{1-p}\mathrm{Cov}_{p,G}[K,{\cal N}].
    \label{eq:avbridges_perc}
\end{equation}
Here, we have used that $\avrcm[\edge] \to p$ as $q\to 1$.  The
l.h.s. being a density, which in particular must be non-negative,
shows that the covariance between the number of components $K$ and
the density of edges $\edge$ is non-positive. This is plausible,
because adding edges can never increase the number of components.
In fact on more mathematical grounds, for the monotone case,
$q\geq 1$, this can even be proven, as it follows from the
Fortuin-Kasteleyn-Ginibre (FKG) inequality \cite{grimmett:book},
applicable to the RC model whenever $q\geq 1$. Applied to $K$ and
$\edge$, the FKG inequality yields $\avrcm[K\edge] \leq \avrcm[K]
\avrcm[\edge]$, which in turn particularly applied to $q=1$ shows
that $\operatorname{Cov}_{p,G}[K \edge] = \avperc[K \edge] - \avperc[K]
\avperc[\edge] \leq 0$. Note that the covariance in
\eqref{eq:avbridges_perc} was analyzed for the percolation model
in Ref.~\cite{deng:06}, however without establishing a connection
to the bridge-density. There, the authors discuss the size
dependence of $\operatorname{Cov}_{p_c,\mathbb{Z}^d_L}[K,\edge]$, where
$\mathbb{Z}^d_L$ is the $d$ dimensional hypercubic lattice
with linear dimension $L$ and periodic boundary conditions. The authors find that the
quantity has a leading finite-size correction that scales as
$L^{1/\nu  - d}$. In Sec.~\ref{sec:fss} we provide an explanation
for this. in terms of the bridge density.

Before we proceed to discussing the behavior on finite lattices in the RC setting, we
derive a symmetry relation for the density of bridges, valid for $\mathbb{Z}^2$,
which for the special case $p=1/2$ implies the previously shown result that
$\mathbb{E}_{1/2,\mathbb{Z}^2}[\mathcal{B}]=1/4$ for critical percolation on
$\mathbb{Z}^2$. Let us naturally view the vertex set of $\mathbb{Z}^2$, denoted by
$V(\mathbb{Z}^2)$, as the set of (ordered) tuples, whose entries are integers, and
the set of edges of $\mathbb{Z}^2$, denoted by $E(\mathbb{Z}^2)$, is the set of pairs
of vertices which differ in precisely one coordinate by $\pm 1$. Then we define a
\textbf{sub-box} of $\mathbb{Z}^2$, denoted by $G_n=(V_n, E_n)$ for $n\geq 1$, where
$V_n = \{(x,y)\in V(\mathbb{Z}^2) \vert -n\leq x,y\leq n \}$ and $E_n =
E(\mathbb{Z}^2)\cap V_n^2$. This allows us to consider a sequence of sub-boxes (in
$n$), that ``approaches'' $\mathbb{Z}^2$. It is not hard to see that each of the
$G_n$'s is planar, hence we can use Euler's formula for planar graphs, stating that
for any $A\subseteq E_n$, $K(A) = \vert V_n\vert - \vert A \vert + F(A) - 1$, where
$F(A)$ is the number of faces in any planar embedding of the graph $(V_n,A)$.  Taking
expectations on both sites and dividing by $\vert E_n\vert$, we obtain
$$
\frac{\mathbb{E}_{p,G_n}[K]}{\vert E_n \vert} = \frac{\vert V_n \vert}{\vert E_n \vert} - p + \frac{\mathbb{E}_{p,G_n}[F]}{\vert E_n \vert}
- \frac{1}{\vert E_n \vert}.
$$
Furthermore it can be verified that for any planar graph the number of components in the primal graph $(V,A)$
equals the number of faces induced by the dual configuration $A^\star$ in the dual graph $(V^\star,A^\star)$, in other words
we have applied to this case, $F_{G_n}(A) = K_{G_n^\star}(A^\star)$, see
e.g. \cite{grimmett:95}. Thus, we can re-write the above 
$$
\frac{\mathbb{E}_{p,G_n}[K]}{\vert E_n \vert} = \frac{\vert V_n \vert}{\vert E_n \vert} - p + \frac{\mathbb{E}_{1-p,G_n^\star}[K]}{\vert E_n \vert}
- \frac{1}{\vert E_n \vert}.
$$
which uses the duality of  bond percolation, i.e., $\mathbb{P}_{p,G}(A) = \mathbb{P}_{1-p,G^\star}(A^\star)$,
valid for any planar graph $G$.
We can now differentiate both sides with respect to $p$ and apply the
Russo-Margulis formula  \eqref{eq:rsfmlperc} to obtain
\[
\mathbb{E}_{p,G_n}[\mathcal{B}] = p - \frac{p}{1-p} \mathbb{E}_{1-p,\mathcal{G}_n^\star}[\mathcal{B}].
\]
Now assuming that $\lim_{n\rightarrow \infty} \mathbb{E}_{p,G_n}[\mathcal{B}]  = \mathbb{E}_{p,\mathbb{Z}^2}[\mathcal{B}]$
and exploiting self-duality of $\mathbb{Z}^2$ in the form of the plausible assumption that
$\lim_{n\rightarrow \infty} \mathbb{E}_{1-p,G_n^\star}[\mathcal{B}] = \mathbb{E}_{1-p,\mathbb{Z}^2}[\mathcal{B}]$,
we obtain
\[
\mathbb{E}_{p,\mathbb{Z}^2}[\mathcal{B}] = p - \frac{p}{1-p} \mathbb{E}_{1-p,\mathbb{Z}^2}[\mathcal{B}].
\]
Finally, for the particular choice of
$p=1/2$ we recover the (asymptotic) result $\mathbb{E}_{1/2,\mathbb{Z}^2}[\bridge] = 1/4$ \cite{xu:14}.
Moreover the above duality result implies a symmetry of $\mathbb{E}_{p,\mathbb{Z}^2}[\bridge]$
around the self-dual or critical point $p=1/2$. In other words determining
$\mathbb{E}_{p,\mathbb{Z}^2}[\bridge]$ on $[0,1/2]$ suffices to obtain
$\mathbb{E}_{p,\mathbb{Z}^2}[\bridge]$ on the entire interval $[0,1]$.

\begin{figure}[tb]
    % generated with inkscape
    \centering
    \includegraphics[width=.6\columnwidth]{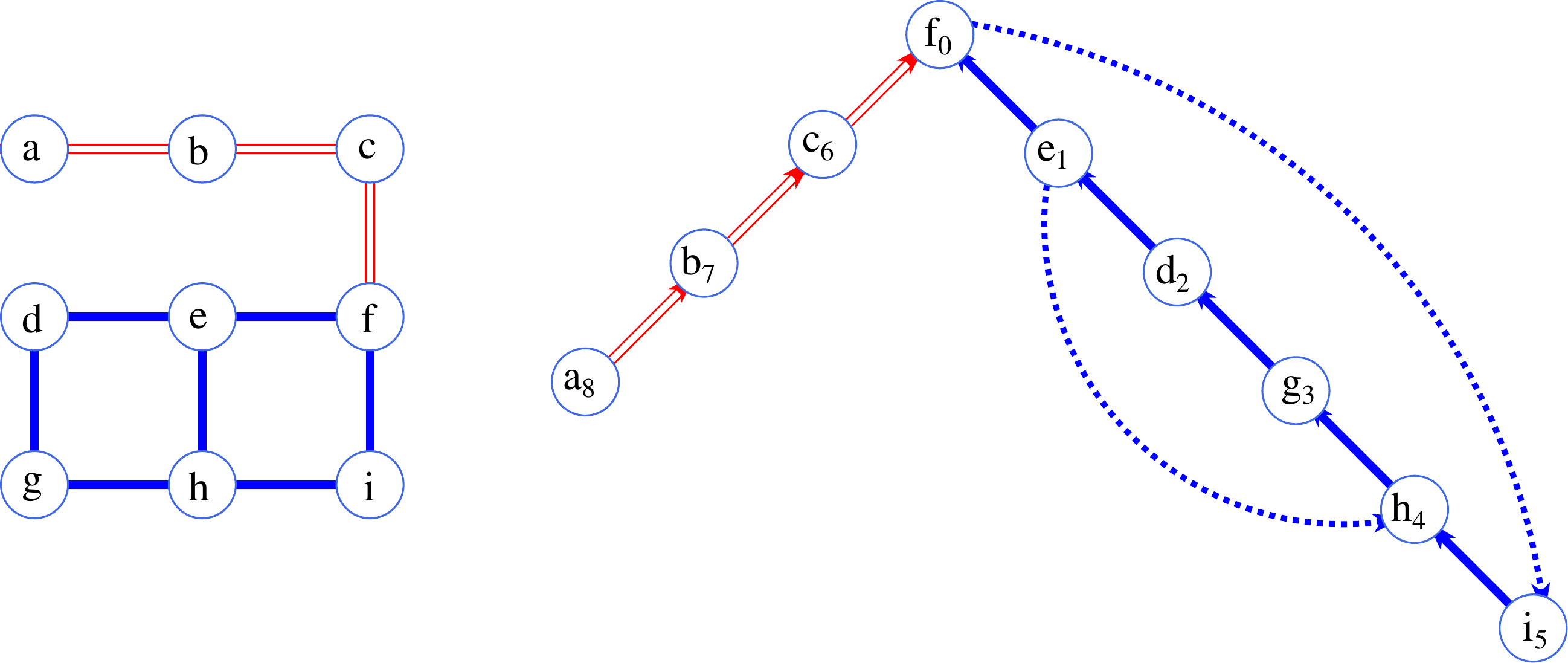}
    \caption{A graph (left panel) with one of its depth-first search trees (right
      panel). Red (hollow) and blue (solid) edges correspond to bridges and non-bridges, respectively.
      The dashed edges in the right panel
      are back-edges. Note that the depth-first search tree together with its
      back-edges can be decomposed into two chains, corresponding to the two
      back-edges, and any non-bridge is in at least one chain. \label{fig:dfsfig}}
\end{figure}

\subsection{Numerical analysis for finite lattices\label{sec:numerical_analysis}}

In order to confirm the asymptotic results (\ref{eq:asymp}) for the square lattice
and to study the finite-size corrections for finite lattices, we performed Monte
Carlo simulations of the RC model on the $L\times L$ square lattice with periodic
boundary conditions, in the range of continuous phase transitions $0< q\le 4$. We
used a recent implementation \cite{elci:13} of (the Metropolis variant of) Sweeny's
algorithm \cite{sweeny:83} for $q<1$ and the Chayes-Machta-Swendsen-Wang algorithm
\cite{chayes:98a} for $q\geq 1$. We determined the number of bridges in a given
configuration by means of the algorithm introduced in Ref.~\cite{schmidt:13}. In
contrast to the approach used in \cite{xu:14,deng:14}, this is a linear time
algorithm applicable to any finite graph. It does not depend on a medial graph
analysis, which facilitates the study of higher dimensional and non-planar
systems. Here we briefly describe the algorithm and refer the reader to
Ref.~\cite{schmidt:13} for more details. The main idea is to construct a depth-first
search (DFS) tree for any component in the spanning subgraph $(V,A)$.  Any edge in
$A$ that is not in a DFS tree is called a \textbf{back-edge} (there are precisely
$\vert A \vert - \vert V \vert + K(A)$ back-edges). Because a back-edge connects
vertices already connected in a DFS tree, it follows that any back-edge is a
non-bridge in $(V,A)$; see Fig.~\ref{fig:dfsfig} for an illustration of this
situation. In order to be able to determine whether a given edge in a DFS tree is a
bridge in $(V,A)$, the algorithm introduces a chain decomposition of the graph, where
for any back-edge there is exactly one chain (ignoring chains consisting only of one
vertex). Now, if a given tree edge is part of such a chain, then it is in at least
one cycle and hence a non-bridge. Finally, a careful construction that avoids the
iteration over overlapping chains ensures that the algorithm classifies all edges
into bridges and non-bridges in linear time.

\begin{figure}[tb]
    % script: figs/bridges_asymptotics.py
    \centering
    \input{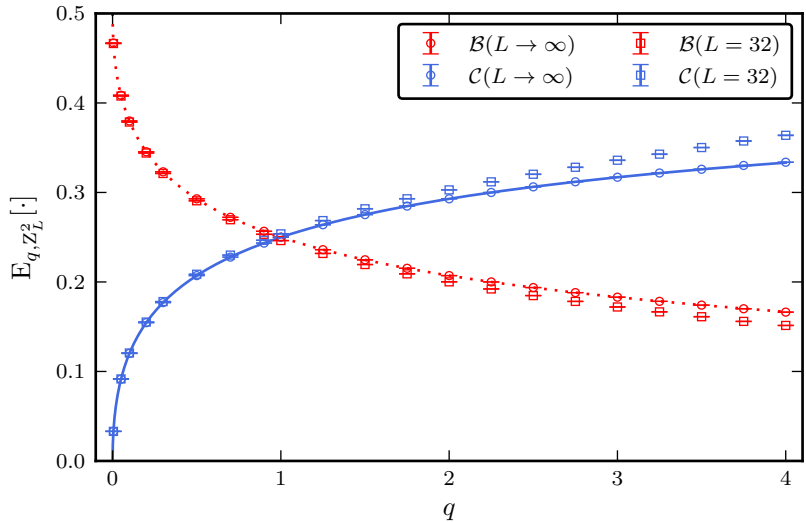}
    \caption{Critical bridge- and non-bridge density for a $L=32$ system
    shown as square symbols. The circles show the asymptotic values
    derived from a finite-size scaling analysis using the ansatz \eqref{eq:bridge_fss}.
    For comparison the solid (blue) and dotted (red) line show the exact asymptotic values
    of non-bridges and bridges, respectively, according to \eqref{eq:asymp}. \label{fig:bridge_nonbridge}}
\end{figure}

%%%%%%%%%%%%%%%%%%%%%%%%%%%%%%%%%%%%%%%%%%%%%%%%%%%%%%%%%%%%%%%%%%%%%%%%%%%%%%%%%%%%%%%%%5
\subsection{Finite-size corrections\label{sec:fss}}

The bridge-edge relation \eqref{eq:avbridges} also allows us to understand the
finite-size corrections to the bridge density. It relates $\avrcm[\bridge]$ to the
density of edges which, in turn, is related to the internal energy via
\eqref{eq:energy_vs_bonds}. Finite-size corrections to the energy density
$u_{\beta,q,G}$ close to a point of a second order phase transition, on the other
hand, are widely studied and well understood from the theory of finite-size
scaling. To formulate it, we need to associate a linear size $L$ to $G$, such that
$|V| = L^d$, as is the case for a lattice graph with spatial dimension $d$. Then, the
standard ansatz for the singular part $f_s$ of the free-energy density $f =
-\log{[Z_{P}(\beta,q,G)]}/(\beta L^d)$ as a function of the reduced temperature
$t=(T-T_c)/T_c$, the external field $h$, and the leading irrelevant field $v$ is
given by \cite{privman:privman}
\begin{equation}
    f_s(t,h,L) = L^{-d} F{\left(tL^{1/\nu},hL^{d-\beta/\nu},vL^{-\theta/\nu} \right)}.
    \label{eq:scaling}
\end{equation}
The internal energy $u_{\beta_c,q,\mathbb{Z}^d_L}$ can be expressed in terms of the
first derivative of $f$ with respect to $t$, evaluated at $t=h=0$. Furthermore it is
plausible to assume \cite{privman:privman,salas:00} that the non-singular part of $f$
has no size-dependence and directly yields the value obtained in the infinite-volume
limit. We hence expect the following scaling of the bridge density:
\begin{equation}
  \avrcmcL[\bridge] = \avrcmcI[\bridge] +  L^{\frac{1}{\nu}-d}\big( b+ \cdots \big).
    \label{eq:bridge_fss}
\end{equation}
Here, we introduced the $L$-subscript to denote the expectation for the model on any
lattice graph with linear dimension $L$, corresponding to the universality class of
the $q$-state Potts model on $\mathbb{Z}^d_L$.  For the case of $\mathbb{Z}^2_L$, the
asymptotic density $\avrcmcI[\bridge]$ is the one of $\mathbb{Z}^2$, as given by
Eq.~(\ref{eq:asymp}). For other 2D lattices such as the triangular and honeycomb
lattice $\avrcmcI[\bridge]$ can be derived in a similar way using well-known
expressions for the critical internal energy there, see, e.g.,
Ref.~\cite{wu:82a}. The scaling form (\ref{eq:bridge_fss}) itself, however, is
universally valid.

\begin{figure}[tb]
    %script ./figs/pseudo_bridges_and_bridges_fss_exponent.py
    \centering
    \input{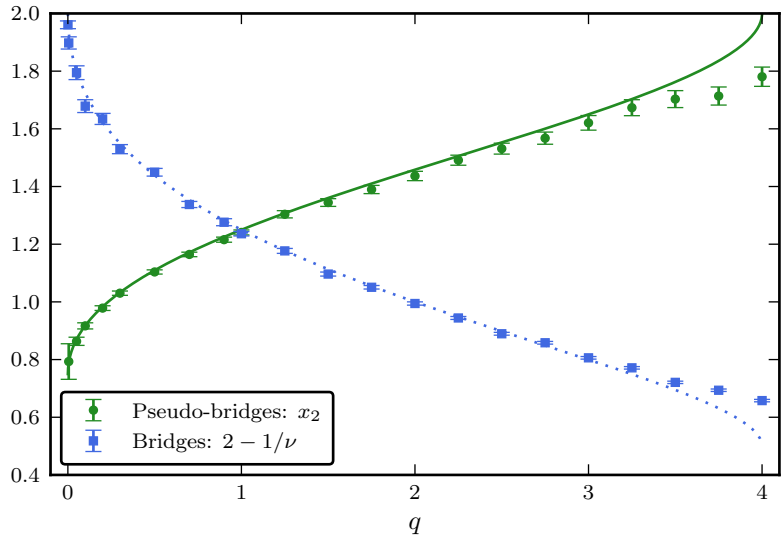}
    %\includegraphics[width=.6\columnwidth]{fss_bridge_density_exponent}
    %\caption{Comparison of numerically extracted exponent of leading finite-size
    %    correction for the density of bridges. The solid line shows
    %the exact Coulomb-gas value.}
    \caption{Comparison of numerically extracted exponents of the leading finite-size
        correction for the density of bridges and pseudo-bridges. The solid and dotted line show
    the exact Coulomb-gas value for pseudo-bridges and bridges, respectively. The deviations for values of $q$ around $4$ are caused
    by strong sub-leading finite-size corrections.}
    \label{fig:bridge_fss}
\end{figure}

We now turn to a more detailed analysis of the corrections for the case of the square
lattice.  We used the above bridge detection algorithm to numerically determine the
size dependence of the bridge density for the critical RC model on $\mathbb{Z}^2_L$
for a number of values of the cluster weight $q$. Testing for scaling of the form
(\ref{eq:bridge_fss}), we fitted the ansatz $a + bL^{c}$ to our Monte-Carlo
estimates, using the method of least squares. Up to cluster weights of about $q
\approx 3.25$, all results were found to be in agreement with the implication of
Eqs.~(\ref{eq:bridge_fss}) and (\ref{eq:asymp}) that $a = 1/[2(1+\sqrt{q})]$ and $c =
1/\nu-d$, where $d=2$. Fixing the parameter $a$ at this exact value, yields more
precise fit results for the exponent $c$. These data are shown in
Fig.~\ref{fig:bridge_fss}, together with the exact values of $1/\nu-d$ known from the
Coulomb gas mapping \cite{nienhuis:domb,coniglio:89}. The deviations observed for
$q\gtrsim 3.25$ are attributed to the presence of higher-order corrections to the
form (\ref{eq:bridge_fss}) which increase in strength on approaching the tricritical
point $q_c = 4$.

Regarding the prefactor $b$ of the leading term $L^{1/\nu - d}$
extracted from the fits, we find that it is negative for the
bridge density, hence the asymptotic result $1/[2(1+\sqrt{q})]$ is
approached from below. The corresponding constant for
$\mathbb{E}_{q,\mathbb{Z}^2_L}[\nbridge]$ is positive such that the limit is approached
from above. This holds for all values analyzed in $q\in[0,4]$.
Thus in illustrative words, the configurations on $\mathbb{Z}^2_L$
still ``feel'' the ``extra'' edges closing the lattice to form a
torus and hence they are typically less ``fragile'' than
corresponding configurations on $\mathbb{Z}^2$. Interestingly,
when one considers the finite-size dependence of $\mathbb{E}_{q,\mathbb{Z}^2_L}[\edge]$
one finds an excess of edges (relative to the asymptotic result
for $\mathbb{Z}^2$) for $q>1$ and a shortfall for $q<1$. Now, due
to the fact that $\avrcm[\edge] = \avrcm[\bridge] +
\avrcm[\nbridge]$, we have that in particular the amplitudes of
the finite-size corrections of $\mathbb{E}_{q,\mathbb{Z}^2_L}[\bridge]$ and
$\mathbb{E}_{q,\mathbb{Z}^2_L}[\nbridge]$ cancel each other for $q=1$, whereas the
modulus of the amplitude for $\mathbb{E}_{q,\mathbb{Z}^2_L}[\bridge]$ is larger
(smaller) than the one for $\mathbb{E}_{q,\mathbb{Z}^2_L}[\nbridge]$ for $q<1$ ($q>1$),
respectively. We omit the details on the size dependence of
$\mathbb{E}_{q,\mathbb{Z}^2_L}[\nbridge]$, as no new mechanism appears and the above
results on $\mathbb{E}_{q,\mathbb{Z}^2_L}[\bridge]$ can be easily adapted to this case.

\begin{figure}[tb]
    %script: ./figs/fit_double.py
    \centering
    \input{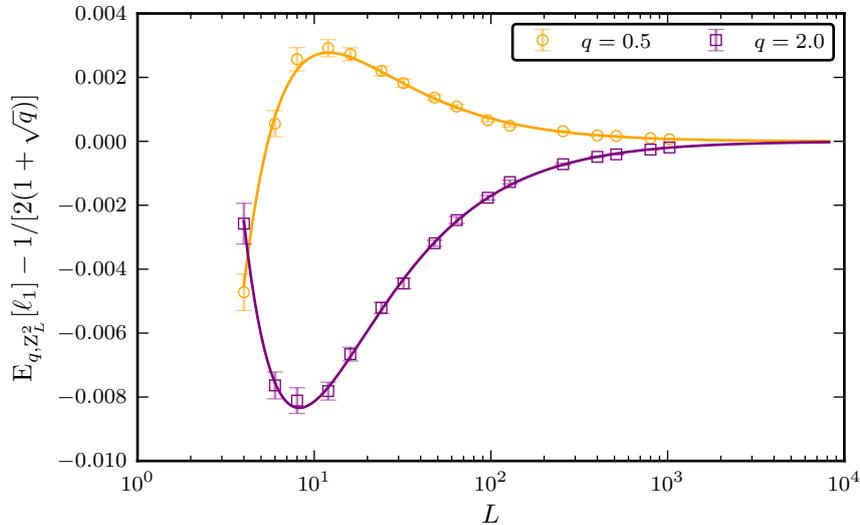}
    \caption{Estimated size dependent deviation $\avrcmcTorus[\ell_1]-1/[2(\sqrt{q} +
      1)]$ for $q=0.5$ and $q=2$ for the self-dual RC model on
      $\mathbb{Z}^2_L$, together with the best fit to the scaling form corresponding
      to (\ref{eq:type1_fss}). All available system sizes starting from $L=4$ were
      included in the two fits and yielded $\chi^2/N_{\rm d.o.f.} = 1.0286$ and
      $0.5554$, for $q=0.5$ and $2$ respectively, where $N_{\rm d.o.f.} = 14$, is the
      number of degrees of freedom for both cluster weights
      considered. \label{fig:fit_type1}}
\end{figure}

Following Xu {\em et al.\/} \cite{xu:14}, we extend our study of finite-size
corrections on the torus $\mathbb{Z}^2_L$ from bridges to the larger class of
``type-1 edges'', that is edges that that have both of their associated loop arcs in
the same loop in the associated loop configuration. While on a planar lattice all
such bonds are bridges, this is not the case for $\mathbb{Z}^2_L$, where also
non-bridges can be of type 1. Such edges are hence called {\bf pseudo-bridges\/}
\cite{xu:14,deng:14}.  If we denote by $L_1(A)$ and $L_2(A)$ the set of (open) edges in
$(V,A)$ that have both of their associated loop arcs in the same loop (type-1 edges)
or in two different loops (type-2 edges), respectively, we have that
\begin{equation}
\mathbb{P}_{p,q,\mathbb{Z}^2_L}[e\in L_1] = \mathbb{P}_{p,q,\mathbb{Z}^2_L}[e\in L_1, e\in B] + \mathbb{P}_{p,q,\mathbb{Z}^2_L}[e\in L_1, e\in C]
                           = \mathbb{P}_{p,q,\mathbb{Z}^2_L}[e\in B] +\mathbb{P}_{p,q,\mathbb{Z}^2_L}[e\in L_1, e\in C] \label{eq:type_1_bridge}.
\end{equation}
The second equality follows from $B\subseteq L_1$, that is any bridge has necessarily
both loop arcs in the same loop (as it cannot enclose any face). Equivalently, we
have $\mathbb{E}_{p,q,\mathbb{Z}^2_L}[\ell_1] = \mathbb{E}_{p,q,\mathbb{Z}^2_L}[\bridge] +
\mathbb{E}_{p,q,\mathbb{Z}^2_L}[\mathcal{P}]$, where {$\mathcal{P}(A)\equiv\left| e\in E : e\in
    L_1(A), e\in C(A) \right|/m$}, i.e., the density of pseudo-bridges in $A$, and
$\ell_1$ denotes the fraction of edges that are of type 1. In
order to understand the general size dependence of
$\avrcmcTorus[\ell_1]$, we determined the number of type-1 edges in
our numerical simulations and fitted a finite scaling ansatz
$\avrcmcTorus[\ell_1] = a + bL^{-\epsilon_1} + cL^{-\epsilon_2}$ to
our Monte-Carlo data. The resulting estimates are consistent with
$a=1/[2(1+\sqrt{q})]$ and $\epsilon_1 = 2-1/\nu $ as well as
$\epsilon_2 = x_2$. In a second step, we also performed fits with
$c$ fixed to its exact value and $\epsilon_1$ and $\epsilon_2$
fixed to the values implied by the Coulomb gas for the
identifications $\epsilon_1 = 2-1/\nu $ and $\epsilon_2 = x_2$,
such that $b$ and $c$ were the only free parameters in a now
linear fit. As it is apparent from the parameters collected in
Table \ref{tab:type_1}, the quality of these fits is very good,
even including the smallest system sizes $L=4$, $8$, $16$, $32$,
strongly suggesting the following asymptotic form for the square
lattice
\begin{equation}
\avrcmcTorus[\ell_1] \sim \avrcmcsq[\bridge] + b L^{1/\nu - d} +
cL^{-x_2}, \label{eq:type1_fss}
\end{equation}
where again $\avrcmcsq[\bridge] = 1/[2(1+\sqrt{q})]$. Figure
\ref{fig:fit_type1} shows our numerical data for the two cluster
weights $0.5$ and $2$, together with the corresponding best fits.

%%%%%%%%%%%%%%%%%%%%%%%%%%%%%%
\begin{table}[tb]
\centering
%  \begin{ruledtabular}
    \begin{tabular}{cccccc}
        $q$ & $b$ & $c$ & $\chi^2/N_{\rm d.o.f.}$ & $L_{\rm min}$ & $Q$ \\ \hline \hline

$3.5$&$ -0.1410(7)$&$ 0.504(7)$&$ 1.0354$&$ 4 $&$ 0.4135$\\
$3.5$&$ -0.1405(8)$&$ 0.49(2) $&$ 0.9722$&$ 8 $&$ 0.4730$\\
$3.5$&$ -0.1406(9)$&$ 0.50(4) $&$ 1.0155$&$ 16$&$ 0.4270$\\
$3.5$&$ -0.140(1) $&$ 0.4(1)  $&$ 1.1116$&$ 32$&$ 0.3514$\\
\hline

$2$&$ -0.218(3)$&$  0.393(7)$&$ 0.5554 $&$  4 $&$ 0.9007$\\
$2$&$ -0.219(3)$&$  0.40(1) $&$ 0.6303 $&$  8 $&$ 0.8183$\\
$2$&$ -0.219(4)$&$  0.40(2) $&$ 0.7086 $&$  16$&$ 0.7173$\\
$2$&$ -0.218(6)$&$  0.38(4) $&$ 0.8504$&$  32$&$ 0.5580$\\
\hline

$1.25$&$ -0.27(1)$&$ 0.32(2)$ &$ 0.8451$&$ 4 $&$  0.6199$ \\
$1.25$&$ -0.26(2)$&$ 0.31(3)$ &$ 0.9229$&$ 8 $&$  0.9229$ \\
$1.25$&$ -0.25(3)$&$ 0.30(4)$ &$ 0.8482$&$ 16$&$  0.5819$ \\
$1.25$&$ -0.31(4)$&$ 0.41(7)$ &$ 0.4951$&$ 32$&$  0.8607$ \\
\hline
$0.9$ & $ -0.21(3)$ & $0.19(3)$ & $0.4983$ &$4$ & $0.9356$ \\
$0.9$ & $ -0.22(5)$ & $0.20(4)$ & $0.5010$ &$8$ & $0.9155$ \\
$0.9$ & $ -0.23(7)$ & $0.21(6)$ & $0.5237$ &$16$ & $0.8748$ \\
$0.9$ & $ -0.3(1) $ & $0.2(9) $ & $0.6113$ &$32$ & $0.7692$ \\
\hline
$0.5  $&$  -0.336(8) $&$  0.192(4)  $&$ 1.0286$ & $4$ & $ 0.4203$ \\
$0.5  $&$  -0.32(1)  $&$  0.188(5)  $&$ 1.0466$ & $8$ & $ 0.4019$ \\
$0.5  $&$  -0.32(2)  $&$  0.188(7)  $&$ 1.2546$ & $16$ &$ 0.2502$\\
$0.5  $&$  -0.33(4)  $&$  0.190(9)  $&$ 1.5533$ &$32$& $ 0.1332$\\
\hline
$0.05 $&$  -0.429(6) $&$  0.033(1)  $&$ 0.7662$ &$4$ &$ 0.7073$\\
$0.05 $&$  -0.41(1)  $&$  0.033(1)  $&$ 0.6955$ &$ 8$ &$ 0.7575$\\
$0.05 $&$  -0.43(3)  $&$  0.033(1)  $&$ 0.6882$ &$ 16$&$ 0.7365$\\
$0.05 $&$  -0.43(7)  $&$  0.033(1)  $&$ 0.8571$ &$ 32$&$ 0.5522$\\
\hline
$0.005$&$  -0.466(5) $&$  0.0040(4) $&$ 0.9548$ &$ 4$&$ 0.4979$\\
$0.005$&$  -0.49(1)  $&$  0.0045(5) $&$  0.6865$ &$8$ &$0.7663$\\
$0.005$&$  -0.49(3)  $&$  0.0046(6) $&$ 0.6857$ &$16$ &$0.7389$\\
$0.005$&$  -0.38(8)  $&$  0.0041(7) $&$  0.5806$ &$32$ & $0.7948$ \\ \hline\hline
    \end{tabular}
\caption{Estimates of the fitting parameters $b$ and $c$ in the
finite-size scaling ansatz \eqref{eq:type1_fss} of the density of
type-1 edges. $Q$ denotes the quality of fit \cite{numrec} or
confidence level, that is the probability that $\chi^2$ would
exceed the empirical value, under the assumption that the imposed
statistical model is correct. In what follows we refer to fits
with confidence level $\geq 10\%$ simply as ``good'' fits.
\label{tab:type_1}}
%  \end{ruledtabular}
\end{table}
%%%%%%%%%%%%%%%%%%%%%%%%%

It is interesting to compare the general form (\ref{eq:type1_fss}) to the results
found in Ref.~\cite{xu:14} for the percolation case, corresponding to $q\to
1$. There, finite-size corrections to the bridge density on the torus
$\mathbb{Z}^2_L$ were found to decay with exponent $-x_2=-5/4$. While this is
numerically consistent with our findings since $1/\nu-d=-x_2$ for percolation
\cite{coniglio:82,duplantier:89,deng:04,deng:10}, it is clear from the present
analysis that the relevant exponent describing finite-size corrections to the bridge
density is $1/\nu-d$. Indeed, the values of $-x_2$ and $1/\nu-d$ strongly differ for
$q\ne 1$, cf.\ Fig.~\ref{fig:bridge_fss}. Moreover, for the square lattice with
periodic boundary conditions Xu {\em et al.\/} \cite{xu:14} showed rigorously that
$\mathbb{P}_{1/2,1,\mathbb{Z}^2_L}[e\in L_1] = \mathbb{E}_{1/2,1,\mathbb{Z}^2_L}[\ell_1] = 1/4$, independent of
$L$. This is consistent with our general form (\ref{eq:type1_fss}) for the
percolation case with $1/\nu-d=-x_2$ if the amplitudes $b$ and $c$ cancel. Indeed,
this is what we find from the fit data in Table \ref{tab:type_1}, which clearly
support the statement that $b=-c$ as $q \to 1$, as required by the rigorous finding
that $\mathbb{E}_{1/2,1,\mathbb{Z}^2_L}[\ell_1] = 1/4$. Hence, it is a particular cancellation of
finite-size effects that leads to the size-independence observed for percolation.
Further, it is known that for $q\rightarrow 0$ with $p=p_{\rm sd}(q)$, one recovers
the uniform spanning tree model, for which clearly no pseudo-bridges exist, that is
one can easily verify that, on $\mathbb{Z}^2_L$, $\avrcmcTorus[\ell_1] \rightarrow \frac{1}{2} -
\frac{1}{2} L^{-2}$ for $q\rightarrow 0$, which is in agreement with
\eqref{eq:type1_fss}, due to $1/\nu \rightarrow 0$ and the numerical observation (see
Table \ref{tab:type_1}) that $a\rightarrow -1/2$ and $b\rightarrow 0$ for
$q\rightarrow 0$. We remark that the fitting estimates for $q$ close to one need to
be treated with caution, as here the two exponents $1/\nu - 2$ and $-x_2$ come very
close to each other, and hence it is numerically very difficult to distinguish the
two contributions and the corresponding constants $a$, $b$. However, outside a
suitable ``safety''-window around $q=1$, it appears that both $a$, $b$ are increasing
with $q$.

%%%%%%%%%%%%%%%%%%%%%%%%%%%

We close the analysis of size-dependent effects by establishing a relationship for
percolation between pseudo-bridges and a particular arm event which shows the
observed $L^{-x_2}$ decay. A pseudo-bridge is a non-bridge that resides on a cross
cluster that winds simultaneously along both directions on the torus
\cite{xu:14}. Furthermore, any pseudo-bridge has both associated loop-arcs in the
medial graph in the same loop. In other words, a pseudo-bridge is a non-bridge on a
cross cluster that is pivotal to the existence of the cross cluster, i.e., if $A$ is
a configuration that contains a cross cluster, then removing a pseudo-bridge $e$
implies that $A_e$ has no remaining cross cluster. Therefore pseudo-bridges are in a
certain sense the ``bridges'' of the cross cluster. We can translate this into a
(polychromatic) four-arm event \cite{beffara:11, bollobas:06} as follows: Any edge,
say $e$, that is pivotal to the cross cluster has the property that starting from $e$
two paths of open edges extend to a distance $L/2$ (here we use the toroidal geometry
and in particular the absence of boundaries). Since $e$ is pivotal for the cross
configuration, there must also be two paths of closed (dual) edges starting at the
dual vertices associated to $e$ and extending to a distance $L/2$, which precisely
ensure that there can be no alternative, $e$-avoiding, path that would preserve the
cross configuration property upon removal of $e$. Thus what we constructed is a
particular instance of an poly-chromatic four-arm event, of alternating primal and
dual ``color'', in the annulus centered around the edge $e$ with outer radius
$L/2$. The probability of such an event decays for large $L$ as $L^{-x_4^{\rm
    {(P)}}}$, where $x_4^{\rm {(P)}}$ is the poly-chromatic four-arm exponent, which
equals $5/4$ for critical percolation\footnote{In fact the value can be rigorously
  established for critical site percolation on the triangular grid, thanks to the
  celebrated work of Smirnov \cite{smirnov:01a}.}. Using the fact that $x_4^{\rm
  {(P)}} = x_2$, where the latter is the two-arm exponent as defined for instance in
\cite{deng:10} and referred to by Xu {\em et al.\/} in
\cite{xu:14} we have the appearance of the exponent $x_2$.

We note that the notion of pseudo-bridges and the corresponding scaling analysis
given above is specific to the case of locally planar lattices. The concentration on
the square lattice was for practical reasons only, however, and we expect the same
results, only with different amplitudes, for the cases of other lattices such as the
triangular or honeycomb cases with periodic boundary conditions.

%%%%%%%%%%%%%%%%%%%%%%%%%%%%%%%%%%%%%%%%%%%%%%%%%%%%%%%%%%%%%%%%%%%%%%%
\section{Other types of pivotal edges\label{sec:candidatebridges}}

So far our analysis has focused on open edges, that is bridges and
non-bridges. Surprisingly, the analysis of closed edges turns out to be very
interesting too, in that it allows us to link several seemingly unrelated quantities,
studied in the literature in different contexts, to the density of bridges. Moreover
the study of closed edges provides a further probe for the cluster structure, for
instance a large density of candidate non-bridges suggests that the clusters are more
likely to self-entangle than to overlap with other clusters. In order to analyze the
corresponding densities consider a given closed edge $e=(x,y)$.  Clearly we have that
when $x\nleftrightarrow y$ then $e$ is a candidate bridge (and vice versa), i.e., a
candidate bridge corresponds to a pair of nearest-neighbor vertices that are not
connected.  This observation allows us to state the following almost trivial but
useful identity that holds for arbitrary $e=(x,y)\in E$:
\begin{equation*}
    \probrcm[e \in \overline{B}]  =  \probrcm[x \nleftrightarrow y].
\end{equation*}
In fact, it is possible to relate the r.h.s.\ above to $\probrcm[e \in B]$, the
probability that $e$ is a bridge. To see this, note that for fixed $e=(x,y)$ any
configuration $A\subseteq E$ that has the property $x\nleftrightarrow y$ can be
related one-to-one to the configuration $A+e$, in which $e$ is a bridge. Furthermore,
as the insertion of $e$ reduces the number of components by one, it follows that
$\probrcm[A] = q(1-p) \probrcm[A^e] / p$ which in turn implies
\begin{equation}
\probrcm[x \nleftrightarrow y] =  \sum_{A\subseteq E} \probrcm[A]
\indicator_{\left\{x \stackrel{A}{\nleftrightarrow} y\right\}}
=   \frac{q(1-p)}{p} \sum_{A\subseteq E} \probrcm[A+e] \indicator_{\left\{e \in B(A+e)\right\}} =\frac{q(1-p)}{p} \probrcm[e \in B].
\label{eq:conbri}
\end{equation}
Hence we have $\probrcm[e \in \overline{B}] = \frac{q(1-p)}{p} \probrcm[e \in B]$,
and therefore using the bridge formula (\ref{eq:avbridges}) and the definition
(\ref{eq:connectivity2}) of the nearest-neighbor connectivity ${\cal E}$ we obtain
the general result
\begin{equation}
    \avrcm[\cbridge] = 1-\avrcm[\mathcal{E}] = \frac{q}{1-q} \left(\frac{\avrcm[\edge]}{p} - 1 \right).
\label{eq:bridgecand}
\end{equation}
The corresponding asymptotic values for the square lattice follow immediately:
\begin{equation}
    \avrcmcsq[\cbridge] = \frac{\sqrt{q}}{2(1+\frac{1}{\sqrt{q}})}, \quad
    \avrcmcsq[\cnbridge] = \frac{1}{2(1+\sqrt{q})}.
    \label{eq:asymp2}
\end{equation}
As a result of the relation to the edge density, we again conclude that the leading
finite-size corrections are given by $L^{1/\nu - d}$. These findings are in line
with those of Hu {\em et
  al.\/} in Ref.~\cite{hu:14}, who find for percolation that for $L\rightarrow
\infty$ on $\mathbb{Z}^2_L$ one has $\mathbb{E}_{p,\mathbb{Z}^2_L}[\mathcal{E}] \rightarrow 3/4$. On the
other hand, it was shown in Ref.~\cite{xu:14} that the asymptotic bridge density for
critical percolation on $\mathbb{Z}^2_L$ is $1/4$ (which also follows from our result
\eqref{eq:avbridges}). These two results are hence clearly consistent with the
relation \eqref{eq:conbri}.  Finally from \eqref{eq:bridgecand} and \eqref{eq:asymp2}
one concludes that
\begin{equation*}
    \avrcmcTorus[\mathcal{E}] \rightarrow
    \frac{2+\sqrt{q}}{2\left(1+\sqrt{q} \right)} \textnormal{ for } L \rightarrow
    \infty,
\end{equation*}
which is consistent with the results of Ref.~\cite{hu:14}.
%%%%%%%%%%%%%%%%%%%%%%%%%%%%%%%%%%%%%%%%%%%%%%%%%%%%%

%%%%%%%%%%%%%%%%%%%%%%%%%%%%%%%%%%%%%%%%%%%%%%%%%%%%
\section{Maximal bridge density for percolation\label{sec:bridge_maximum}}

\begin{figure}[tb]
  \centering
  % script: figs/plot_max_frag_perc.py
  \input{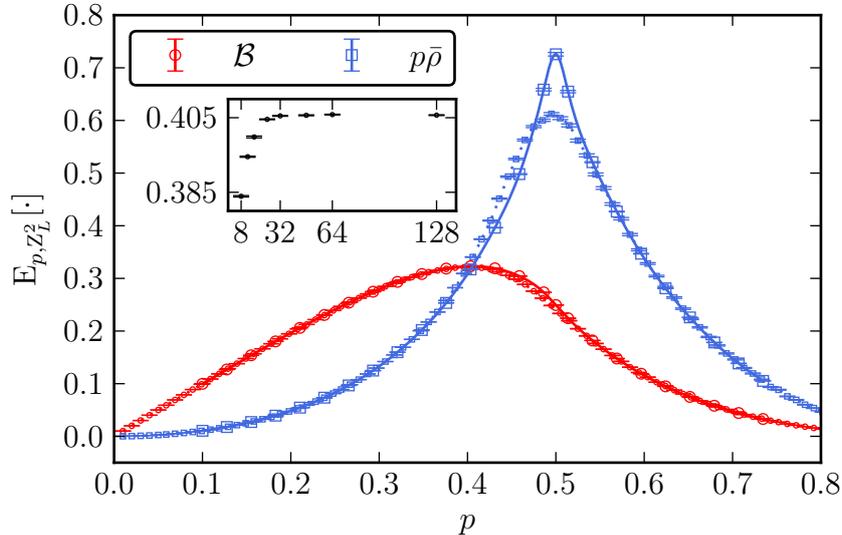}
  \caption{Numerically estimated density of bridges and bridge load for percolation
    on $\mathbb{Z}^2_{L}$ for $L=16$ and $L=128$, corresponding to the smaller and larger markers, respectively. 
    Additionally, the inset shows our estimates for $p_f$ for system sizes from $L=8$ up to $L=128$.
    The estimates for the location of the crossing ``saturate'' to a value  $0.4056(5)$, and therefore we
    conclude that $p_{\rm f}=0.4056(5)$ for $\mathbb{Z}^2$, according to \eqref{eq:cond}. We point out that we
    used a naive depth-first search traversal based algorithm for determining the bridge load, so we could not study
    systems with $L > 128$ in practise. The lines in the figure are obtained by interpolating
    the data points using splines.}
  \label{fig:maxperc}
\end{figure}

It is apparent from Fig.~\ref{fig:bridge_nonbridge_ising} that the
expected bridge density $\mathbb{E}_{p,q,\mathbb{Z}^2_L}[\bridge]$ is maximal for $p_{\rm
f}(q) < p_{\rm sd}(q)$ for any $q\in(0,4]$ \cite{elci:15}. The
reason for this is a priori not clear and here we focus on the
percolation case $q=1$ and obtain an expression for $\frac{d}{dp}
\avperc[\bridge]$, that is for general graphs $G$, to determine its zero(s) one of which must
correspond to the location of the sought maximum.  In principle we
could again apply the Russo-Margulis formula directly to $\vert B
\vert$ to obtain the required derivative. However, there is an
alternative, more geometric, approach. In Ref.~\cite{coniglio:82},
Coniglio showed that the $p$-derivative of $\probperc[x
\leftrightarrow y]$ for two vertices $x,y\in V$ (not necessarily
neighbors) satisfies the following differential equation:
\begin{equation}
\frac{\rm d}{{\rm d}p} \probperc[x \leftrightarrow y] = \frac{1}{p} \avperc[\lambda_{x,y}],
\label{eq:coniglio}
\end{equation}
where $\lambda_{x,y}(A)$ is the number of bridges on any\footnote{It can be verified
  that the number of bridges on a self-avoiding path between to vertices is an
  invariant, hence it does not matter which self-avoiding path one chooses (in case
  more than one exists).} self-avoiding path in $A$ connecting $x$ and $y$.  We
emphasize that for a given edge $(x,y)\in E$ the quantity $\lambda_{x,y}$ can be
non-zero, even if $(x,y)\notin B(A)$. This is because $\lambda_{x,y}$ counts the
number of bridges in the spanning subgraph $(V,A)$ that would disconnect $x$ from $y$
upon removal. As an example, consider vertices $b$ and $e$ in the left panel in
Figure \ref{fig:dfsfig}. Even with $(b,e)$ not being part of the considered spanning
subgraph, we have $\lambda_{b,e} = 2$, because edges (in fact bridges) $(b,c)$ and
$(c,f)$ are essential for the connectivity of $b$ and $e$. Note that bridge $(a,b)$
is not essential for the connectivity of $b$ and $e$ as its removal does not
disconnect $b$ and $e$.
%We note
%that this quantity is $0$ if there exists no path and hence $x$ and $y$ are
%disconnected, or if $x$ and $y$ are in a cycle and hence connected by at least two
%edge-disjoint paths.
Note that relationship \eqref{eq:coniglio} also follows straightforwardly from
the Russo-Margulis formalism applied to the indicator function of the event $x\leftrightarrow y$.
As a result of the bijection relation (\ref{eq:conbri}) we can transform
\eqref{eq:coniglio} into a differential equation for the bridge density:
\begin{equation}
\frac{\rm d}{{\rm d}p} \avperc[\bridge] = -\frac{1}{(1-p)\vert E \vert} \sum_{(x,y)\in E} \avperc[\lambda_{x,y}] +
\frac{1}{p(1-p)} \avperc[\bridge].
\end{equation}
We emphasize that now $\lambda_{x,y}$ is only evaluated for
nearest-neighbor pairs, that is for $x,y\in V$ with $(x,y)\in E$.
Observe
\begin{equation}
\sum_{(x,y)\in E} \lambda_{x,y}(A) = \sum_{(x,y)\in E} \sum_{e\in E}  \mathds{1}_{\left\{x \stackrel{A}{\leftrightarrow} y\right\}}
\mathds{1}_{\left\{x\stackrel{A_e}{\nleftrightarrow} y\right\}}  =
\sum_{e\in E} \left[ \sum_{(x,y)\in E}  \mathds{1}_{\left\{x \stackrel{A}{\leftrightarrow} y\right\}}
\mathds{1}_{\left\{x\stackrel{A_e}{\nleftrightarrow} y\right\}} \right].
\label{eq:rho_def}
\end{equation}
Now, for a given edge $e\in E$ define $\rho_e(A)$ as the sum in
the square brackets in \eqref{eq:rho_def} or, equivalently, to be
$0$ when $e$ is not a bridge in $A$ and otherwise the number of
nearest-neighbor pairs for which the bridge $e$ lies on any (all)
self-avoiding path(s) in $(V,A)$ between them. In other words
$\rho_e(A)$ counts for how many nearest-neighbor pairs edge $e$
contributes to $\lambda_{x,y}$. We call $\rho_e$ the
\textbf{bridge load} of $e$. Note that we have $\rho_e(A)\geq
\mathds{1}_{\{e\in B(A)\}}$ and hence $\mathbb{E}_{p,q}[\rho_e]
\geq \mathbb{P}_{p,q}[e\in B]$. For an example consider the left
panel in Figure \ref{fig:dfsfig}, here edges $(a,b), (b,c), (c,f)$
have bridge loads $2$, $3$, $3$, respectively.  It is evident that
in a certain sense $\rho_e$ generalizes the bridge density and
encodes how relevant a given bridge is for the connectivity at
large. We emphasize that one only has to count connected
\emph{neighbor} pairs.  We finally obtain:
\begin{equation}
\frac{\rm d}{{\rm d}p} \avperc[\bridge] = -\frac{1}{(1-p)} \avperc[\bar{\rho}] +
\frac{1}{p(1-p)} \avperc[\bridge],
\label{eq:deriv}
\end{equation}
where we defined $\bar{\rho}(A) \equiv \frac{1}{\vert E \vert} \sum_{e\in E}
\rho_e(A)$. Therefore we find that the l.h.s. in \eqref{eq:deriv} vanishes for
$p=p_{\rm f}$, where $p_{\rm f}$ is the solution of the following equation
\begin{equation}
\mathbb{E}_{p_{\rm f},G}[\bridge] = p_{\rm f} \mathbb{E}_{p_{\rm f},G}[\bar{\rho}].
\label{eq:cond}
\end{equation}
We remark that $p_{\rm f}$ is a graph dependent quantity and we suppress the explicit
dependence.  In Fig.~\ref{fig:maxperc} we illustrate this condition for a small
system size and the case $G=\mathbb{Z}^2_L$.  Moreover, under translational
invariance (which is given, e.g., for $\mathbb{Z}^2_L$), \eqref{eq:cond} is
equivalent to $\mathbb{P}_{p_{\rm f},G}[e \in B] = p_{\rm f} \mathbb{E}_{p_{\rm
    f},G}[\rho_e]$ for arbitrary $e\in E$. This has a nice intuitive interpretation:
$\mathbb{E}_{p,G}[\rho_e]$ counts the typical number of nearest-neighbor pairs at the
interface between two clusters that are glued together \emph{only} by the presence of
bridge $e$.  There are two effects at work here: increasing $p$ typically increases
the size of the clusters and therefore possibly also the number of nearest neighbors
on the interface. Additionally, however, increasing $p$ will eventually create
additional links between the two clusters, in which case $e$ ceases to be a bridge,
such that this effect decreases $\probperc[e\in B]$.  Now, the maximal fragility
(maximum of $\probperc[e \in B]$) is attained for the bond density $p_{\rm f}$, where
the expected number of closed edges between two clusters glued together by $e$ equals
precisely the probability that $e$ is a bridge. Increasing $p$ further strengthens
the cohesion between the clusters, and hence the probability of $e$ being a bridge
starts to decrease afterwards.

In what follows
we restrict the analysis to the case $G=\mathbb{Z}^2_L$ in order
to extract the asymptotic value of $p_f$ for $\mathbb{Z}^2$.
Consider Figure \ref{fig:maxperc}, where we numerically confirm
condition \eqref{eq:cond} and estimate $p_{\rm f} =0.4056(5)$ for
$\mathbb{Z}^2$. We note that, albeit the maximum of
$\mathbb{P}_{p,\mathbb{Z}^2}[e\in B]$ is attained for $p=p_{\rm f}$, the bridge load
$\mathbb{E}_{p,\mathbb{Z}^2}[\rho_e]$ increases further and is apparently maximized
for $p=p_{\rm sd}(1) = p_c = 1/2$.  This nicely reflects the
intricate cluster structure at criticality. One might wonder how
the bridge density can decrease but the overlap continues to
increase. This can happen when in most (probable) configurations
$e$ is not a bridge, but for configurations where $e$ is a bridge
one has typically a very large self-entanglement of the clusters
attached to the two ends of $e$, outweighing the effective
decrease of $\mathbb{P}_{p,\mathbb{Z}^2_L}[e \in B]$. This effect is most drastic for
$p=p_c$. We note that Coulomb gas arguments predict that the
l.h.s.\ of \eqref{eq:deriv} remains finite at $p=p_{\rm sd}(1) =
1/2$, which clearly implies that $\mathbb{E}_{p,\mathbb{Z}^2}[\rho_e]$ remains finite
at $p=1/2$. More precisely we can show that the peak of $\mathbb{E}_{q,\mathbb{Z}^2_L}[\rho_e]$
scales as $L^{2-2x_2}$ for $q\in[0,4]$, where in particular $x_2 = 5/4$ for percolation [see the
discussion in the next section and in particular
\eqref{eq:bridge_load_expr}]. Unfortunately, so far we have not
been able to establish an exact value for $p_{\rm f}(1)$ let alone in
the correlated setting $q\neq 1$ for $\mathbb{Z}^2$. This is a
challenging problem we wish to address in a forthcoming study.

%\showthe\columnwidth

%%%%%%%%%%%%%%%%%%%%%%%%%%%%%%%%%%%%%%%%%%%%%%%%%%%%%%%%%%%%%%%%
\section{Bridge fluctuations\label{sec:bridge_correlations}}

As we have seen, the bridge-edge formula (\ref{eq:avbridges})
together with the theory of finite-size scaling, provides a rather
complete understanding of the critical bridge density.  Higher
moments of the bridge distribution can be discussed with similar
techniques as we will show now for the example of the variance.
Naively, one might expect that this variance
$\operatorname{Var}_{p,q,G}[\vert B\vert ]$ only depends on the
fluctuations of $N$. In this case the critical variance would
follow $\operatorname{Var}_{q,L}[\vert B \vert]/L^d
\approx L^{\alpha/\nu}$, which, in two dimensions, in turn would
imply a divergence with $L$ for $q\geq 2$ and a ``saturation'' for
$q<2$ \cite{deng:10}. As we will show however, the story is not
quite as simple.

To work out the fluctuations of the bridge density, we apply the
Russo-Margulis formula to the second derivative of the partition
function $Z_{p,q,G}\equiv Z_{\rm RC}(p,q,G)$, which we then in turn equate to the expression
one obtains by explicit differentiation. To start with, we have,
using \eqref{eq:der_freeenergy_2},
\begin{align}
\partial_{p}^2 \partfrcm &= (1-q) \partial_p \left( \frac{\partfrcm}{p} \avrcm[\vert B \vert] \right) \nonumber\\
%   &= (1-q) \partial_p \left( \frac{\mathbb{E}_p\left[q^K \vert B \vert\right]}{p} \right),   \nonumber\\
   &=  -\partfrcm \frac{1-q}{p^2} \avrcm\left[\vert B \vert \right] +
   \frac{1-q}{p} \partial_p \avperc\left[q^K \vert B \vert \right]. \nonumber
\end{align}
Let us now focus on the second term:
\begin{align}
\frac{1-q}{p} \partial_p \avperc\left[q^K \vert B \vert \right]
% &= \frac{1-q}{p} \sum_{e\in E} \mathbb{E}_{p}\left[\delta_e\left(q^K \vert B \vert \right)\right], \nonumber \\
&= \frac{1-q}{p} \sum_{e,f\in E} \avperc\left[\delta_e\left(q^K \mathds{1}_{\{f \in B\}} \right)\right] \nonumber\\
&= \frac{1-q}{p^2} \sum_{e,f\in E} \sum_{A\subseteq E:\atop e \in A} \probperc[A] \left[q^{K(A)} \mathds{1}_{\{f\in B(A)\}}
-q^{K(A_e)} \mathds{1}_{\{f\in B(A_e)\}}\right] \label{eq:fluct_intermediate}.
\end{align}
We can split the inner sum into two sums corresponding to $e\in B(A)$ and $e\in
C(A)$, of which we first consider the former
\begin{equation*}
\frac{1-q}{p^2} \sum_{e,f\in E} \sum_{A\subseteq E:\atop e \in B(A)} \probperc[A]
\left[\mathds{1}_{\{e=f\}} q^{K(A)} \mathds{1}_{\{f\in B(A)\}}  +
\mathds{1}_{\{e\neq f\}}q^{K(A)} (1-q)  \mathds{1}_{\{f\in B(A)\}}\right].
\end{equation*}
A few comments are in order: If $e=f$ we clearly have that
$e=f\notin B(A_e)$ hence we recover the expected bridge density.
For $e\neq f$ it is important to observe that removing a bridge
$e$ cannot influence the pivotality of an occupied edge $f$, in
particular when $f\in B(A)$ then also $f\in B(A_e)$, whenever $e$
is a bridge in $A$. The above can therefore be re-written as
\begin{equation*}
\frac{1-q}{p^2} \partfrcm  \avrcm\left[ \vert B \vert \right] +
\frac{(1-q)^2}{p^2} \partfrcm \sum_{e\neq f\in E}   \probrcm[e\in B, f\in B].
\end{equation*}
For the non-bridge sum of (\ref{eq:fluct_intermediate}) note that only summands with
$e\neq f$ contribute, because otherwise if $e\in C(A)$ then trivially $f=e\notin
B(A)$, $B(A_e)$. Furthermore, if $e\neq f$ such that $e\in C(A)$ and $f\in B(A)$ then
we have also a vanishing contribution because deleting a non-bridge cannot change the
fact that $f$ is a bridge. Thus, there can only be a contribution for a configuration
$A$ such that $e\in C(A)$ and $f\in C(A)$ as well as both edges are in one cycle and
deleting $e$ will destroy the cycle and hence cast $f$ into a bridge in
$A_e$. Moreover, both edges $e$ and $f$ must only be in one cycle that has no
edge overlap with other cycles (imagine two clusters glued together in parallel by the edges $e$ and $f$,
hence in particular there are no additional links between the two clusters). Write $e
\overset{A}{\Leftrightarrow} f$ for the above event involving edge $e$ and $f$. We
obtain for the second term:
\begin{equation*}
-\partfrcm \frac{1-q}{p^2}\sum_{e\neq f\in E} \probrcm\left[e \Leftrightarrow f \right].
\end{equation*}
We therefore obtain eventually for $\partial_p^2 Z_{p,q}$:
\begin{align}
\partial_p^2 \partfrcm &= \partfrcm \frac{1-q}{p^2} \left(
(1-q)\sum_{e\neq f\in E}   \probrcm[e\in B, f\in B] -
\sum_{e\neq f \in E} \probrcm[e \Leftrightarrow f] \right) \nonumber\\
&= \partfrcm \frac{1-q}{p^2} \left( (1-q) \operatorname{Var}_{p,q,G}[\vert B \vert ]
 + (1-q) \avrcm[\vert B \vert]^2 - (1-q)\avrcm[\vert B \vert]
 - \sum_{e\neq f \in E} \probrcm[e \Leftrightarrow f] \right). \nonumber
\end{align}
On the other hand one can show by explicit differentiation and after some straightforward
but tedious algebra that one has:
$$
\frac{1}{\partfrcm} \partial_p^2 \partfrcm= \frac{\avrcm[N]
\left(2p - 1 -2p\vert E \vert \right) - \vert E \vert p^2 + \vert
E \vert^2 p^2 + \operatorname{Var}_{p,q,G}[N] +
\avrcm[N]^2}{p^2(1-p)^2}.
$$
Equating both expressions for $\partial_p^2 \partfrcm$ with subsequent rearranging yields
\begin{align}
\operatorname{Var}_{p,q,G}[\vert B \vert] &=
\frac{\avrcm[N] \left(2p - 1 -2p\vert E \vert \right) - \vert E \vert p^2 + \operatorname{Var}_{p,q,G}[N] +
 \left(\avrcm[N] - \vert E \vert p\right)^2 + 2\vert E \vert p\avrcm[N] }{(1-q)^2(1-p)^2}  \nonumber \\
 &- \avrcm[\vert B\vert ]^2 + \avrcm[\vert B\vert ] + \frac{1}{1-q}
\sum_{e\neq f\in E}   \probrcm[e \Leftrightarrow f].
\nonumber
\end{align}
Using the bridge edge identity \eqref{eq:avbridges} we can simplify the above
to
\begin{equation}
\frac{\operatorname{Var}_{p,q,G}[\vert B \vert]}{\vert E \vert } =
\frac{\avrcm[\mathcal{N}] \left(2p - 1 \right) - p^2 +
  \operatorname{Var}_{p,q,G}[N]/\vert E \vert  }{(1-q)^2(1-p)^2}  +
\avrcm[\mathcal{B}] + \frac{1}{1-q} \frac{1}{\vert E \vert }
\sum_{e\neq f\in E}   \probrcm[e \Leftrightarrow f] \label{eq:var_b_rslt}.
\end{equation}
We emphasize that \eqref{eq:var_b_rslt} is an exact result valid
for any $p$ and $q$ as well as any graph $G$, thus it has the same
status as the bridge-edge identity \eqref{eq:avbridges}.
Furthermore, we remark that for $e=(u,v)$ and $f=(x,y)$ we have by
a bijection argument, similar to the one used in
Sec.~\ref{sec:candidatebridges} to derive the relationship between
the bridge density and nearest-neighbor connectivity:
$$
\probrcm[e\Leftrightarrow f] = \frac{p^2}{(1-p)^2} \frac{1}{q}
 \probrcm[(u,v) \veebar (x,y)] .
$$
Here $\probrcm[(u,v) \veebar (x,y)]$ for $(u,v),(x,y)\in
E$ is the probability that the two nearest-neighbor pairs belong
to two different clusters, such that the two distinct clusters
each contain one vertex from $\{u,v\}$ and one vertex from
$\{x,y\}$ (see Fig.~$1(c)$ in Ref.~\cite{vasseur:12}).

We pause for a moment, and remark that completely analogous arguments can be used to
obtain an expression for the $p$-derivative of $\avrcm[\mathcal{B}]$, studied
in Section \ref{sec:bridge_maximum}, in terms of the probabilities of events
$e\Leftrightarrow f$.  A consequence of this is that we can obtain an explicit
expression for $\avrcm[\bar{\rho}]$:
\begin{equation}
\avrcm[\bar{\rho}] = \avrcm[\mathcal{B}] + \frac{1}{\vert E \vert}
\frac{1-p}{p} \sum_{e\neq f\in E} \probrcm[ e \Leftrightarrow f].
\label{eq:bridge_load_expr}
\end{equation}
In particular, this shows explicitly that the bridge load can increase further beyond
the point $p_{\rm f}(G)$, for which $\avperc[\mathcal{B}]$ is maximal. Moreover,
as we show below, we find that at criticality, in two dimensions, $\mathbb{E}_{q,\mathbb{Z}^2_L}[\bar{\rho}] \approx
L^{2-2x_2}$, which together with $x_2= 5/4$ for the case $q=1$ in two dimensions,
shows that the peak at $p=1/2$ in Figure \ref{fig:maxperc} stays bounded as
$L\rightarrow \infty$.

Coming back to the study of fluctuations of $\vert B\vert$, we
focus in what follows on the self-dual line for the RC model on
$\mathbb{Z}^2_L$ and study the continuum limit. As shown by
Vasseur {\em et al.\/} \cite{vasseur:12} one has for two pairs of
neighboring vertices $(u,v)$ and $(x,y)$ at distance $r$ the
following asymptotic for large $r$
$$
\probrcmcsq[(u,v) \veebar (x,y)] \sim A(q) r^{-2x_2(q)}.
$$
This follows from the construction of a four-leg watermelon event,
due to the four hulls propagating from the neighborhood of $(u,v)$
to the neighborhood of $(x,y)$, which are associated to the two
clusters involved. We hence expect the following asymptotic
behavior
\begin{align}
  \frac{1}{\vert E \vert}\sum_{e\neq f\in E} \probrcmcTorus[e \Leftrightarrow f] &=
  \frac{1}{\vert E \vert} \sum_{(u,v) \neq (x,y) \in E } \probrcmcTorus[(u,v) \veebar (x,y)]  \nonumber \\
  &\approx A(q) \int_1^{\frac{L}{2}} {\rm dr} 2 \pi r r^{-2x_2(q)} \nonumber \\
  &= \alpha(q) L^{2-2x_2(q)} + \beta(q),
\end{align}
where $\alpha(q)$ and $\beta(q)$ are two $q$ dependent constants.

\begin{figure}[tb]
    \centering
    %script ./figs/plot_fluc_exponents.py
    %\the\textwidth
    \input{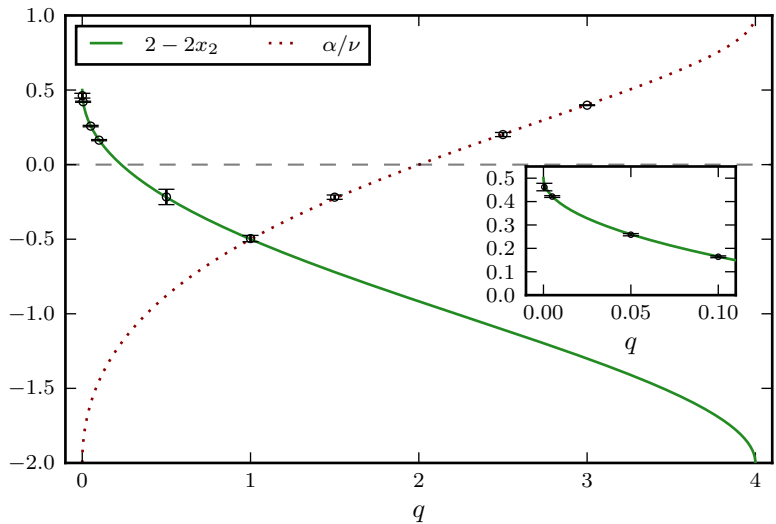}
    \caption{The exponents $\alpha/\nu$ and $2-2x_2$ appearing in the system size scaling
    of $\operatorname{Var}_{q}[\vert B \vert]/m$ at criticality. The solid and dotted lines show the
    exact value of $2-2x_2$ and $\alpha/\nu$, respectively, following from the Coulomb gas mapping. The symbols denote our
    numerical estimates from fitting to the variance of the bridge density, cf.\ the
    data collected in Table \ref{tab:fitting_nikos}.
 \label{fig:exponents}}
\end{figure}

Inspecting the form (\ref{eq:var_b_rslt}), we hence see that the
fluctuation of the bridge density has two contributions, one
related to the variance of the edge density that scales as
$L^{\alpha/\nu}$ at criticality, and another one related to the
above mentioned watermelon event with scaling proportional to
$L^{2-2x_2}$. In order to decide which contribution constitutes
the dominant leading scaling, it is useful to recall some relevant
exact results from the Coulomb gas approach to two dimensional
critical phenomena \cite{nienhuis:domb}.  One finds that
$\alpha/\nu = 4 - 12/g$ and $x_2 = 2 - (4-g)(4+3g)/8g$, where for
critical 2D RC models one has to relate the Coulomb gas coupling
$g$ to the cluster weight $q$ via $q = 2 + 2 \cos{(g\pi/2)}$
\cite{deng:10}.  A small calculation shows then that
$L^{\alpha/\nu}$ constitutes the leading contribution to the
system size scaling of the variance whenever $q \ge 1$. In
particular, for $q > 2$, where $\alpha/\nu > 0$, this translates
into a divergence of $\operatorname{Var}_{q,\mathbb{Z}^2_L}[\vert B
\vert]/L^2$ with $L$, resembling the specific-heat singularity in
the same cluster weight regime. On the other hand, for $q < 1$ the
leading term is proportional to $L^{2-2x_2}$.  Interestingly (and
somewhat unexpected), one can explicitly verify that for $q<
\tilde{q}=4\cos^2{(\pi/\sqrt{3})}=0.2315891\cdots$ the variance of
$\vert B \vert$ diverges with the exponent $2-2x_2>0$. This
particular value $\tilde{q}$ follows from the condition
$2-2x_2(\tilde{q}) = 0 \Leftrightarrow x_2(\tilde{q}) = 1$, which,
using the above expressions, yields a quadratic equation for
$\tilde{g}=g(\tilde{q})$ (or $\tilde{q}$): $8\tilde{g} =
(4-\tilde{g})(4+3\tilde{g})$, with the two solutions
$\tilde{g}=\pm 4/\sqrt{3}$. Taking only the positive solution we
find that $\tilde{q}=2+2\cos{(2\pi/\sqrt{3})} =
4\cos{(\pi/\sqrt{3})}^2$.  The situation is summarized in the
dotted and solid lines of Fig.~\ref{fig:exponents}, showing the
Coulomb gas values of $\alpha/\nu$ and $2-2x_2$, respectively. We
also analyzed the variance of the bridge density numerically. The
fitting functions and resulting parameters are summarized in Table
\ref{tab:fitting_nikos}, and the corresponding parameter estimates
are indicated by the symbols in Fig.~\ref{fig:exponents}. Clearly,
we find excellent agreement with the expectations from
Eq.~(\ref{eq:var_b_rslt}) discussed above. For the marginal value
$\tilde{q} = 0.2315891\cdots$ we expect a logarithmic divergence,
and we indeed find the corresponding form to yield the best fit to
our simulation data.

\renewcommand{\arraystretch}{1.3}

\begin{table}[tb]
\centering
\begin{tabular}{cccccccc}
\hline \hline
$q$ & $L_{\min}$ &  $\chi^2/N_{\rm d.o.f.}$ & Model & $\correxp$ & $\alpha/\nu$ & $2-2x_2$ \\
\hline
$0.0005$ & $12$ & $1.01$  & $a+bL^\correxp$ & $0.462(16)$ & $-1.9576$ & $\mathbf{0.4752}$\\
$0.005$ & $4$ & $1.15$ & $a+bL^\correxp$ & $0.422(3)$ & $-1.8679$ & $\mathbf{0.4222}$\\
$0.05$ & $16$ & $0.58$ & $a+bL^\correxp$ & $0.258(5)$ & $-1.6005$ & $\mathbf{0.2599}$\\
$0.1$ & $8$ & $0.76$ & $a+bL^\correxp$ & $0.164(4)$ & $-1.4492$ & $\mathbf{0.1648}$\\
$0.231589$ & $8$ & $1.01$ & $a+b\log{(L)}$ & $-$ & $-1.1962$ & $0$\\
$0.5$ & $12$ & $1.4$ & $a+bL^\correxp + cL^{\alpha/\nu}$ & $-0.217(51)$ & $-0.8778$ & $\mathbf{-0.2191}$\\
$1$ & $6$ & $0.7$ & $a+bL^\correxp\left[1 + c \log{(L)} \right]$ & $-0.496(21)$ & $-0.5$ & $-0.5$\\
$1.5$ & $4$ & $1.05$ & $a+bL^\correxp + cL^{2-2x_2}$ & $-0.218(15)$ & $\mathbf{-0.2266}$ & $-0.7205$\\
$2.5$ & $48$ & $0.94$ & $a+bL^\correxp$ & $0.202(14)$ & $\mathbf{0.2036}$ & $-1.1052$\\
$3$ & $6$ & $1.14$ & $a+bL^\correxp$ & $0.398(3)$ & $\mathbf{0.4}$ & $-1.3$\\
\hline
\hline
\end{tabular}
\caption{Numerical results for the leading exponent $\kappa$ in the
finite-size scaling of $\operatorname{Var}_{q,\mathbb{Z}^2_L}[\vert B \vert]$.
The two rightmost columns show the exact values obtained from the
Coulomb gas mapping. The exponent values in bold face indicate the asymptotically
dominant behavior. For the cluster weights $q=1.5$ and
$q=0.5$ we also performed a fit to the form $a+bL^\correxp$, which
yielded slightly worse results, due to the proximity of the two
(negative) exponents. In all cases the quality-of-fit $Q$ was at least 5\%.
\label{tab:fitting_nikos}}
\end{table}

%%%%%%%%%%%%%%%%%%%%%%%%%%%%%%%

It remains to discuss the percolation case $q=1$ where the
bridge-edge identity (\ref{eq:avbridges}) becomes singular and the
above derivation needs to be revisited.  Here, we derive the
singular behavior based on another bijection argument that allows
us to harness recent results on logarithmic observables emerging
from a careful analysis of the appropriate logarithmic conformal
field theory description of critical percolation \cite{vasseur:12,
hu:14}. Before we do so, note that in order to extract the
asymptotic scaling of the variance it suffices to study the
(co-)variance $\eta_{e,f} \equiv \probrcm[e \in B , f \in B] -
\probrcm[e \in B] \probrcm[f \in B]$, which relates to
$\operatorname{Var}_{p,q,G}[\vert B\vert]$ via the well known
identity
$$
\operatorname{Var}_{p,q,G}[\vert B \vert] = \sum_{e,f\in E} \eta_{e,f}.
$$
To start with, recall that we consider critical bond percolation
on $\mathbb{Z}^2_L$, i.e. $q=1$ and $p=1/2$ and write
$\mathbb{P}[\cdot]$ for $\mathbb{P}_{1/2,1,\mathbb{Z}^2_L}[\cdot]$. Furthermore,
note that $\mathbb{Z}^2_L$ is a transitive graph, and hence none
of the following events depend on the explicit edge or vertex,
used in the arguments.  Now, fix two edges
$e=(x_1,y_1),f=(x_2,y_2)$ that are distance $r \ll L$ apart.
Consider the event $\{ e\in B \land f \in B\}$. All configurations
contributing to this event can be further sub-divided into two
events, depending on whether $e$ and $f$ belong to the same
connected component in $(V,A)$ or not. Denote the two events by
$\Omega_1$ and $\Omega_2$, respectively.  Choose a configuration
$A$ that belongs to $\Omega_2$, i.e. the two edges $e$ and $f$ are
bridges in $(V,A)$ and belong to two different connected
components. The crucial point is that we can relate $A$ one-to-one
to $A-\{e,f\}$, a configuration where $x_1$, $y_1$, $x_2$, $y_2$
belong to four different clusters. Denote all configurations in
which the four vertices belong to four different components by
$\tilde{\Omega}_2$ (this is a event).  Due to the choice of $q=1$
and $p=1/2$ we have that $\prob{\Omega_2} =
\prob{\tilde{\Omega}_2}$.  Let us now consider the event
$\Omega_1$, i.e. the set of configurations for which $e$ and $f$
belong to the same component and both $e$ and $f$ are bridges.
Now, because both edges are pivotal we can relate any such
configuration $A\in \Omega_1$ one-to-one to a configuration where
$x_1$ and $y_1$ as well as $x_2$ and $y_2$ are disconnected, and
for which $x_1,y_1,x_2,y_2$ belong to three different clusters, of
which one cluster contains one vertex of $\{x_1,y_1\}$ and one of
$\{x_2,y_2\}$.  Denote the corresponding event $\tilde{\Omega}_1$.
Note that any configuration $A\in \Omega_1$, where $e$ and $f$ are
bridges belonging to the same cluster, must yield $3$ disconnected
clusters in $A'=A-\{e,f\}$. This is because the alternative case
of $2$ disconnected clusters in $A'$ would imply that $e$ and $f$
are in a cycle in $A$, which is obviously a contradiction. As
before we have $\prob{\Omega_1} = \prob{\tilde{\Omega}_1}$.
Finally, the probabilities $\prob{\tilde{\Omega}_1}$ and
$\prob{\tilde{\Omega}_2}$ were studied\footnote{Vasseur {\em et
al.\/} in \cite{vasseur:12} denote
  $\mathbb{P}[\tilde{\Omega}_1]$ by $\mathbb{P}_1(r)$ and
  $\mathbb{P}[\tilde{\Omega}_2]$ by $\mathbb{P}_0(r)$.} in \cite{vasseur:12} in the
framework of a corresponding logarithmic conformal field theory.
We note that in the continuum limit both probabilities
$\prob{\Omega_1}$ and $\prob{\Omega_2}$ only depend, to leading
order, on $r$. Since we have $\prob{e\in B, f\in B} =
\prob{\Omega_1} + \prob{\Omega_2}$ we obtain by falling back to
\cite{vasseur:12}:
\begin{equation}
\eta_{e,f} \sim \left[ a + b\log{(r)} \right]r^{-2x_2}
\Rightarrow \frac{\operatorname{Var}_{1/2,1,\mathbb{Z}^2_L}[\vert B \vert]}{L^2} \sim a'+\left[b' + c' \log{(L)}\right] L^{2-2x_2} ,
\label{eq:corr}
\end{equation}
where $x_2 = 5/4$ is the two-arm exponent for critical percolation and $a$, $b$,
$a'$, $b'$ and $c'$ are constants. Our numerical analysis confirmed this scaling, as shown
in Table \ref{tab:fitting_nikos} and Figure \ref{fig:exponents}. Thus considering the
variance of bridges for critical percolation yields yet another manifestation of the
underlying logarithmic conformal field theory \cite{vasseur:12}. We remark that the
same logarithmic multiplicative corrections are expected in higher dimensions
\cite{vasseur:14}. More precisely, one expects even the same scaling form, that is
$$
\frac{\operatorname{Var}_{p_c,1,\mathbb{Z}^d_L}[\vert B\vert]}{L^d} \sim a'' + [b'' + c'' \log{(L)}] L^{2/\nu -d},
$$
where we used the well known fact that $x_2 = d - 1/\nu$ holds for critical
percolation \cite{coniglio:82}. It would be interesting to to verify this theoretical
prediction numerically.

Before we proceed to the next section we briefly discuss the finite-size scaling of
$\operatorname{Var}_{q,\mathbb{Z}^2_L}[\vert C\vert]/|E|$.  This can be worked out by
observing that $\vert C \vert = \vert A \vert - \vert B \vert$ which in turn implies
the general identity
$$
\operatorname{Var}_{p,q,G}[\vert C \vert] = \operatorname{Var}_{p,q,G}[\vert A \vert]
+ \operatorname{Var}_{p,q,G}[\vert B \vert] - 2 \operatorname{Cov}_{p,q,G}[\vert A \vert, \vert B \vert].
$$
Additionally, one can verify that  (see \cite{grimmett:book} or by explicit differentiation of
$Z_{p,q}$)
$$
\operatorname{Cov}_{p,q,G}[\vert A \vert, \vert B \vert] =
p(1-p)\partial_p \avrcm[\vert B \vert].$$ An immediate
consequence of these observations is that
$\operatorname{Var}_{q,\mathbb{Z}^d_L}[\vert C \vert]$ is governed by the same
finite-size scaling as $\operatorname{Var}_{q,\mathbb{Z}^d_L}[\vert B \vert]$ and
we therefore omit further details.

\section{Bridge-free clusters\label{sec:bridge_free}}

In concluding the present study, we return to the discussion of the structure of the
incipient percolating cluster outlined above in the introduction. In the
`links--nodes--blobs' picture of Ref.~\cite{stanley:77} the attention is focused on
the backbone of the cluster as the skeleton ensuring long-range connectivity. Bridges
on the backbone are singled out as the bonds carrying \emph{full} current for the
case of an external potential difference being applied, hence their name 'red
bonds'. Due to the importance of these structures for a number of applications of
percolation theory, significant effort has been invested in the determination of the
associated fractal dimensions $d_\mathrm{BB}$ and $d_\mathrm{RB}$ of the backbone and
the red bonds, respectively \cite{grassberger:99a}. Even in two dimensions, where
most critical exponents and fractal dimensions, including $d_\mathrm{F}$ and
$d_\mathrm{RB}$, are known exactly from Coulomb gas and further arguments
\cite{nienhuis:domb}, there is no exact expression for the fractal dimension of the
backbone.

Clearly, the notions of backbone and red bonds appear to depend on
the concept of long-range conductivity, and it is not obvious how
they could be defined in terms of the more local connectivity
structure of bridges and non-bridges. The concepts of junctions
and bridges introduced in Ref.~\cite{xu:14} are related, but not
commensurate to the idea of separating bridges in the backbone
(red bonds) and bridges in the blobs. We expect that, away from
$q\to 0$, most bridges in the backbone are junctions as the
`links--nodes--blobs' picture tells us that the backbone has
cycles. Still, there are junctions in the dangling ends and there
are branches in the backbone. Also, as Xu {\em et al.\/}
established for percolation in Ref.~\cite{xu:14}, branches and
junctions are asymptotically finite fractions of the bridge set
and hence of the edge set. It is therefore interesting to study
the fractal dimensions of the related edge classes. Removal of
branches from critical configurations will typically only shave
off the last part of dangling ends, and we do not expect this to
alter the fractal dimension of the incipient percolating cluster.
This is indeed what Xu {\em et al.\/} report \cite{xu:14} and we
find the same to be true also for the RC model. On the other hand,
removing the junctions means deleting most of the red bonds and
hence breaks down the percolating cluster into individual blobs.
Their size scales with the backbone fractal dimension as $\propto
L^{d_\mathrm{BB}}$. The same holds true for the case of {\bf
bridge-free\/} clusters where both, branches and junctions, have
been removed. Studying these sets hence provides an alternative
route to the determination of $d_\mathrm{BB}$.

\begin{figure}[tb]
\centering
%script figs/backbone_plot.py
\input{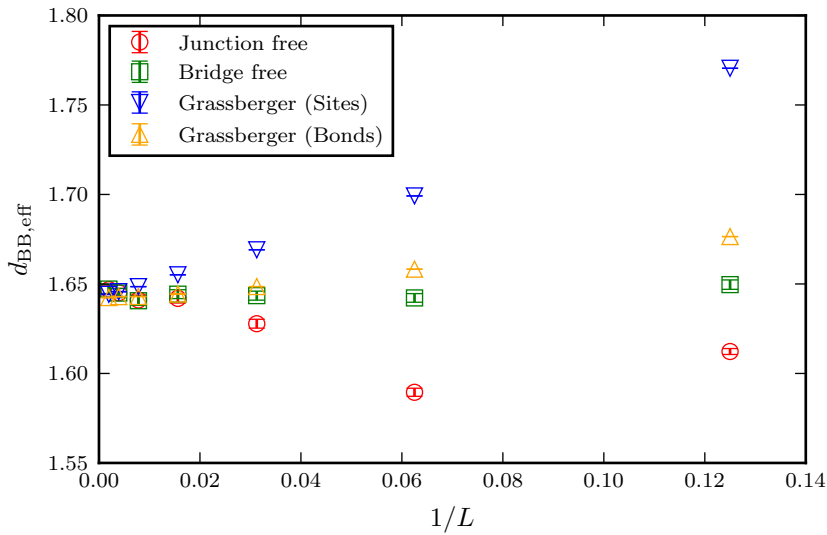}
\caption{Effective backbone scaling exponents according to
  Eq.~(\ref{eq:effective_dimension}) for the $q=1$ RC model extracted from the
  scaling of the backbone sites, the backbone bonds (data from
  Ref.~\cite{grassberger:99a}), the junction-free clusters and the bridge-free
  clusters (our data), respectively.
\label{fig:grassberger}}
\end{figure}

To investigate the utility of this approach for the determination of $d_\mathrm{BB}$,
we determined the scaling of the junction-free and bridge-free clusters for the
percolation case $q=1$.  If we consider the effective, system-size dependent fractal
dimensions \cite{grassberger:99a}
\begin{equation}
  \label{eq:effective_dimension}
  d_{BB,\mathrm{eff}} \equiv \frac{\log[N(2L)/N(L/2)]}{\log 4},
\end{equation}
where $N(L)$ denotes the number of sites in the junction-free or bridge-free clusters
at size $L$, respectively, we can easily compare corrections to scaling for the
different approaches. This is illustrated in Fig.~\ref{fig:grassberger}, comparing
the exponents extracted from the scaling of the junction-free and bridge-free
clusters to the effective exponents found from the scaling of the backbone itself
\cite{grassberger:99a}. Clearly, the scaling of bridge-free clusters is significantly
less affected by scaling corrections than the scaling of the backbone
itself. Compared to this, junction-free clusters show slightly increased corrections,
however interestingly with a correction amplitude of the opposite sign.  We remark
that our estimate for percolation $d_\mathrm{BB} = 1.643(1)$ is consistent with
existing literature values, such as $d_\mathrm{BB} = 1.64336(1)$, $1.6434(2)$,
$1.6432(8)$ in \cite{xu:14, deng:04, grassberger:99a}, respectively. As this is a
small-scale study not using the particularly efficient algorithms available for the
uncorrelated percolation case, the reduced scaling corrections do not immediately
lead to an improved estimate. Instead, we focus on the estimation of $d_\mathrm{BB}$
for a wide range of $q$ values for which no estimates have been previously reported
\cite{deng:04}.  Whereas for the determination of $d_\mathrm{BB}$ directly from the
backbone corrections to scaling need to be carefully taken into account, it turns out
that the bridge-free clusters provide a scaling route less encumbered with such
problems. As is evident from the results collected in Table \ref{tab:backbone}, over
the full range of $q$ values we achieve excellent fits without the inclusion of
scaling corrections.

\begin{table}[tb]
\centering
\caption{Fit results for backbone fractal dimension $d_{BB}$. \label{tab:backbone}}
\begin{tabular}{ccccccc}
\hline \hline
$q$ & $d_\mathrm{BB}$ & $L_{\rm min}$ & $N_{\rm d.o.f.}$ & $\chi^2/N_{\rm d.o.f.}$ & $Q$ \\
\hline
$0.0005$ & $ 1.264(2)$ & $32 $ &$ 8 $ & $  1.0631$ &  $  0.3858 $ \\
$0.0050$ & $ 1.301(1)$ & $16 $ &$ 10$ & $  0.2934$ &  $  0.9829 $ \\
$0.0500$ & $ 1.383(1)$ & $16 $ &$ 10$ & $  0.4661$ &  $  0.9126 $ \\
$0.1000$ & $ 1.425(1)$ & $12 $ &$ 11$ & $  1.1773$ &  $  0.2966 $ \\
$0.2000$ & $ 1.480(1)$ & $12 $ &$ 11$ & $  1.0047$ &  $  0.4390 $ \\
$0.3000$ & $ 1.516(1)$ & $32 $ &$ 8 $ & $  0.7543$ &  $  0.6434 $ \\
$0.5000$ & $ 1.567(1)$ & $24 $ &$ 9 $ & $  0.9153$ &  $  0.5104 $ \\
$0.7000$ & $ 1.603(1)$ & $32 $ &$ 8 $ & $  0.6676$ &  $  0.7206 $ \\
$0.9000$ & $ 1.630(1)$ & $24 $ &$ 9 $ & $  1.0700$ &  $  0.3813 $ \\
$1.0000$ & $ 1.643(1)$ & $24 $ &$ 9 $ & $  1.3227$ &  $  0.2187 $ \\
$1.2500$ & $ 1.670(1)$ & $24 $ &$ 9 $ & $  1.0177$ &  $  0.4227 $ \\
$1.5000$ & $ 1.692(1)$ & $24 $ &$ 9 $ & $  1.1271$ &  $  0.3389 $ \\
$1.7500$ & $ 1.716(2)$ & $128$ &$ 4 $ & $  0.7496$ &  $  0.5581 $ \\
$2.0000$ & $ 1.732(1)$ & $96$ &$ 5 $ & $  1.2095$ &  $  0.3016 $ \\
$2.2500$ & $ 1.744(1)$ & $48 $ &$ 7 $ & $  1.0578$ &  $  0.3880 $ \\
$2.5000$ & $ 1.756(1)$ & $96 $ &$ 5 $ & $  0.5469$ &  $  0.7408 $ \\
$2.7500$ & $ 1.774(1)$ & $96 $ &$ 5 $ & $  1.1302$ &  $  0.3417 $ \\
$3.0000$ & $ 1.785(1)$ & $64 $ &$ 6 $ & $  0.6563$ &  $  0.6851 $ \\
$3.2500$ & $ 1.797(1)$ & $96 $ &$ 5 $ & $  0.9530$ &  $  0.4452 $ \\
$3.5000$ & $ 1.806(1)$ & $96 $ &$ 5 $ & $  0.8288$ &  $  0.5289 $ \\
$3.7500$ & $ 1.828(4)$ & $400$ &$ 2 $ & $  1.0609  $ &  $  0.3462 $ \\
$4.0000$ & $ 1.83(1) $ & $32 $ &$ 7 $ & $  1.3921  $ &  $ 0.2035 $ \\
\hline
\hline
\end{tabular}
\end{table}

\begin{figure}[tb]
\centering
%script ./figs/pub_exponent_bridge_free_backbone.py
\input{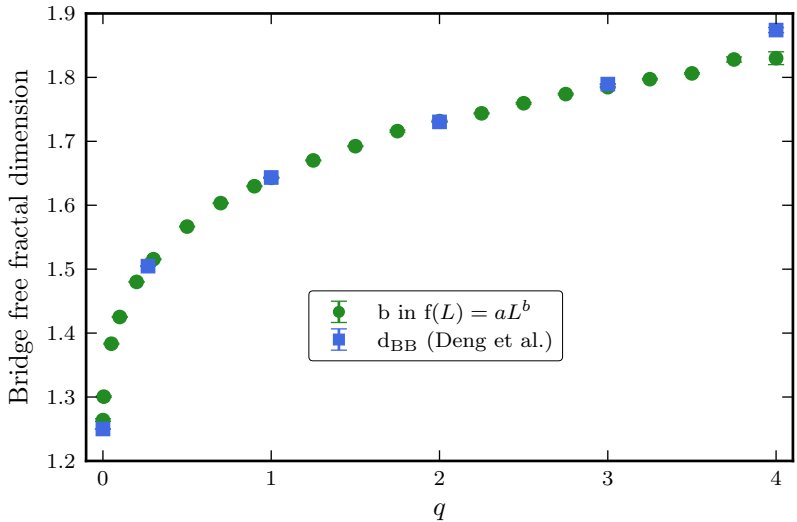}
\caption{Estimated backbone fractal dimension using the method described in
  \cite{xu:14}. For comparison also values from \cite{deng:04} are shown. The
  disagreement for $q=4$ is most likely due to known logarithmic
  corrections.\label{fig:bridge_free}}
\end{figure}

\section{Discussion and outlook\label{sec:conclusion}}

As we have seen, the natural classification of edges into bridges
and non-bridges provides an understanding of how the cluster
weight influences the connectivity in the random-cluster model.
For any $q\ne 1$ we established that both bridges and non-bridges
have expected densities that are linearly related to the overall
density of open edges, such that there is always a non-zero
fraction of both edge types. The derivation is based on an
application of the Russo-Margulis formula to the partition
function of the model, a connection that allows to express
analytical derivatives in terms of combinatorial and geometric
quantities. For the percolation limit $q\rightarrow 1$ the linear
relation does not hold and rather transforms into an intuitive
identity relating the bridge density to the covariance between the
number of connected components and open edges. The latter identity
explains previously observed finite-size corrections for this
covariance in the percolation model \cite{deng:06}. Thus, the
established bridge-edge formula allows us to connect the
finite-size corrections of the densities of bridges and
non-bridges to corrections in the energy density governed by the
exponent $d-1/\nu$. The lack of finite-size effects in the density
of the so-called type-1 edges for the special case of a 2D lattice
with periodic boundary conditions, corresponding to a torus graph,
can be understood from a cancellation of correction terms, and we
clarified the character and origin of correction terms with
exponents $d-1/\nu$ and $-x_2$, respectively \cite{xu:14}.

By studying the dependence of the bridge density on the bond occupation $p$ we
observe that its maximum is attained for $p_{\rm f}(q)$ strictly below the
corresponding critical point $p_{\rm sd}(q)$. We explained this observation for the
percolation case by studying the newly introduced bridge load of an edge. For
percolation on $\mathbb{Z}^2_L$ we estimate $p_{\rm f}=0.4056(5)$. Our numerical
study of this quantity was limited by the computational cost of determining the
bridge load for each edge, which we were not able to calculate with linear
(worst-case) computational effort for all edges. However, we remark that it is
possible to express a condition for $p_{\rm f}$ in terms of easily computable
covariances and moments (in linear time), which however requires to estimate the
difference of quantities of order $O(L^d)$ to a precision of order $O(1)$, which
implies a statistical inefficiency. The construction of a more efficient algorithm
for the determination of the bridge load constitutes an interesting open problem.

Having established the expectation values of the bridge and non-bridge densities, we
turned to a study of the fluctuations.  Again using the Russo-Margulis formalism we
established a second bridge-edge identity relating the variances
$\operatorname{Var}_{p,q,G}[\vert B \vert]$ and $\operatorname{Var}_{p,q,G}[\vert C
\vert]$ to the variance of the bond density and an extra term relating to a specific
four-leg watermelon event \cite{vasseur:12}. For the special case of the
two-dimensional model, we use results on the scaling of this event derived from
conformal field theory \cite{vasseur:12} to predict a singularity in both variances
for $q\leq\tilde{q}\equiv 4 \cos^2{(\pi/\sqrt{3})}=0.2315891\cdots$, an effect
completely absent in the fluctuations of the overall number of open edges.  Numerical
simulations confirm these predictions to a high precision.  The limiting percolation
case required a separate analysis, and here we are able to show that the finite-size
corrections to the density of bridges for critical percolation have logarithmic
multiplicative corrections, a direct consequence of the underlying logarithmic
conformal field theory. It would be interesting to study these fluctuations for
higher dimensional systems to see whether such a singularity appears generically for
some $q<1$, and how this threshold value depends on dimensionality. It follows from
hyperscaling, $\nu d = 2 -\alpha$, and the relation $1/\nu = d_R = d - x_2$
\cite{coniglio:89} valid only for percolation ($q=1$) that in this case $\alpha/\nu =
d-2x_2$, irrespective of dimension $d$. Hence, $q=1$ is a marginal case in all
dimensions below the upper critical one, yielding a multiplicative logarithm in the
variance of the number of bridges. It would be interesting to see whether $d-2x_2 >
\alpha/\nu$ for $q < 1$ in 3D also (see \cite{deng:10} and \cite{vasseur:14} for two
definitions of $x_2$ that can be generalized to $d\geq 3$), thus hinting at an
interesting physical/geometrical difference between the (anti-monotone) $q<1$ and
(monotone) $q>1$ regimes.

Removing bridges leads to fragmentation of clusters as discussed recently in
Ref.~\cite{elci:15}. Removing {\em all\/} bridges decomposes the percolating cluster
into blobs, and it turns out that this approach allows for a very precise
determination of the backbone fractal dimension, much less affected by finite-size
corrections than the more traditional approach of studying the backbone directly
\cite{grassberger:99a}. We provide estimates of this important dimension, which is
one of the few exponents of the random-cluster model in two dimensions for which no
exact expression is known, for a wide range of values in $0< q < 4$, many of which
have been studied here for the first time.

An even more detailed understanding of the role of bridges could be expected from a
generalization of the random-cluster model giving different weights to bridge- and
non-bridge bonds. This problem of a generalized bridge percolation \cite{araujo:11a}
is the subject of a forthcoming study. The $O(n)$ loop model is another general class
of frequently studied systems \cite{domany:81}. As these are graph polynomials as
well (namely over a restricted set of spanning subgraphs, i.e., all loop
configurations or Eulerian subgraphs \cite{deng:07b}), it is therefore tempting to
investigate in how far a similar Russo-Margulis approach can be applied there or,
more generally, to Eulerian subgraph models \cite{deng:07b}, possibly yielding novel
identities.

%%%%%%%%%%%%%%%%%%%%%%%%%%%%%%%%%%%%%%%%%%%%%%%%%%%%%%%%%%%%%%%%%%%%%%%%%%%%%%%
%%%%%%%%%%%%%%%%%%%%%%%%%%%%%%%%%%%%%%%%%%%%%%%%%%%%%%%%%%%%%%%%%%%%%%%%%%%%%%%
\section{Acknowledgement}

We would like to thank Timothy M. Garoni and Youjin Deng for
valuable discussions. E.M.E. is grateful to the members of the
School of Mathematical Sciences at Monash University for their
warm hospitality during his stay in summer 2014, where part of
this work was done. The authors acknowledge funding from the EC
FP7 Programme (PIRSES-GA-2013-612707).

%%%%%%%%%%%%%%%%%%%%%%%%%%%%%%%%%%%%%%%%%%%%%%%%%%%%%%%%%%%%%%%%%%%%%%%%%%%%%%%
%%%%%%%%%%%%%%%%%%%%%%%%%%%%%%%%%%%%%%%%%%%%%%%%%%%%%%%%%%%%%%%%%%%%%%%%%%%%%%%
\appendix
\section*{Appendices}
\addcontentsline{toc}{section}{Appendices}
\renewcommand{\thesubsection}{\Alph{subsection}}
%%%%%%%%%%%%%%%%%%%%%%%%%%%%%%%%%%%%%%%%%%%%%%%%%%%%%%%%%%%%%%%%%%%%%%%%%%%%%%%
%%%%%%%%%%%%%%%%%%%%%%%%%%%%%%%%%%%%%%%%%%%%%%%%%%%%%%%%%%%%%%%%%%%%%%%%%%%%%%%

\subsection{Influence of an edge \label{ap:a}}

In order to prove (\ref{eq:inflbridge}), we first decompose the sum over subgraphs
$A\subseteq E$ into two contributions, corresponding to configurations for which
$e\in A$ and the complementary set of configurations with $e\notin A$.  Then observe
that for $A$ such that $e\in A$ we clearly have $A^{e} = A$ and similarly for $A$
with $e\notin A$ we have $A_{e} = A$. Furthermore we can also relate the
probabilities of $A$ and the modified set $A_e$ or $A^e$ once we know whether $e\in
A$. Using these rules, we arrive at
\begin{eqnarray*}
    \avperc[\delta_e X] &=& \sum_{A \subseteq E} \probperc[A] \lbrace X(A^e) - X(A_e)
\rbrace \\
&=& \sum_{A \subseteq E: \atop e \in A} \probperc[A] \lbrace X(A) - X(A_e) \rbrace +  \sum_{A \subseteq E: \atop e \notin A} \probperc[A] \lbrace
 X(A^e) - X(A_e) \rbrace\\
 &=& \sum_{ A \subseteq E : \atop  e\in A} \probperc[A] \lbrace X(A) - X(A_e) \rbrace +
 \frac{1-p}{p}\sum_{A \subseteq E : \atop e \in A } \probperc[A] \lbrace
 X(A) - X(A_e) \rbrace) \\
 &=& \frac{1}{p} \sum_{A \subseteq E : \atop  e \in A } \probperc[A] \lbrace X(A) - X(A_e) \rbrace.
\end{eqnarray*}
We remark that in the last sum one can safely discard the condition $e\in A$, because 
for $A\subset E$ with $e\notin A$ one trivially has $A_e = A$ and hence $X(A) - X(A_e) = 0$.
%%%%%%%%%%%%%%%%%%%%%%%%%%%%%%%%%%%%%%%%%%%%%%%%%%%%%%%%%%%%%%%%%%%%%%%%%%%%%%%
%%%%%%%%%%%%%%%%%%%%%%%%%%%%%%%%%%%%%%%%%%%%%%%%%%%%%%%%%%%%%%%%%%%%%%%%%%%%%%%

\subsection{Free energy derivative \label{ap:b}}

In order to evaluate $\avperc[q^K]$, note that $K$ only changes on removing a bond
$e$ if it is a bridge, i.e., $K(A) - K(A_e)=-\indicator_{\left\{ e \in B(A) \right\}}$. Hence, we
have
\[q^{K(A)} - q^{K(A_e)}=q^{K(A)}(1-q)\indicator_{\left\{e\in B(A)\right\}}.\]
We can now apply (\ref{eq:inflbridge}) to obtain:
\begin{eqnarray*}
    \avperc[\delta_e q^K] &=& \frac{1}{p} \sum_{A\subseteq E : \atop e \in A} \probperc[A]
    \left[ q^{K(A)} - q^{K(A_e)} \right] \\
     &=& \frac{1-q}{p} \sum_{A\subseteq E : \atop e \in A} \probperc[A]
     q^{K(A)}\indicator_{\left\{e\in B(A)\right\}} \\
     &=& \frac{1-q}{p} Z_\mathrm{RC}(p,q,G) \avrcm[\indicator_{\{e \in B\}}] \\
     &=& \frac{1-q}{p} \avperc[q^K] \probrcm[e \in B].
\end{eqnarray*}
Hence we obtain
\begin{eqnarray*}
\frac{\partial}{\partial p} \log{Z_{\rm RC}(p,q,G)} &=& \frac{\partial}{\partial p} \log \avperc[q^K] \\
&=& \frac{1}{\avperc[q^K]} \frac{\partial}{\partial p} \avperc[q^K] \\
&=& \frac{1}{\avperc[q^K]} \sum_{e \in E} \avperc[\delta_e q^K] \\
&=& \frac{1-q}{p}   \sum_{e \in E} \probrcm[e \in B] \\
&=& \vert E \vert \frac{1-q}{p}  \avrcm[\bridge].
\end{eqnarray*}

\bibliographystyle{elsarticle-num}
%\bibliography{citeulike_nourl_noissn}

\end{document}

%% file: format.tex
\usepackage{amsmath,amssymb,amstext,amsthm,enumerate}

\theoremstyle{plain}

\theoremstyle{definition}

\theoremstyle{remark}

\newcommand{\indicator}{\mathds{1}}
\newcommand{\avrcmc}{\mathbb{E}_{q,G}}
\newcommand{\avrcmcsq}{\mathbb{E}_{q, \mathbb{Z}^2}}
\newcommand{\avrcmcL}{\mathbb{E}_{q, L}}
\newcommand{\avrcmcTorus}{\mathbb{E}_{q, \mathbb{Z}^2_L}}
\newcommand{\avrcmcI}{\mathbb{E}_{q, \infty}}
\newcommand{\avrcm}{\mathbb{E}_{p,q,G}}
\newcommand{\avrcmsq}{\mathbb{E}_{p,q,\mathbb{Z}^2_L}}
\newcommand{\avperc}{\mathbb{E}_{p,G}}

\newcommand{\edge}{\mathcal{N}}
\newcommand{\bridge}{\mathcal{B}}
\newcommand{\nbridge}{\mathcal{C}}
\newcommand{\cbridge}{\overline{\mathcal{B}}}
\newcommand{\cnbridge}{\overline{\mathcal{C}}}
\newcommand{\prob}[1]{\mathbb{P}{\left[{#1}\right]}}

\newcommand{\probrcmcsq}{\mathbb{P}_{q,\mathbb{Z}^2}}
\newcommand{\probrcmcTorus}{\mathbb{P}_{q,\mathbb{Z}^2_L}}

\newcommand{\probrcm}{\mathbb{P}_{p,q,G}}
\newcommand{\probperc}{\mathbb{P}_{p,G}}
\newcommand{\partfrcm}{Z_{p,q,G}}

%% file: paper_arxiv.bbl
\begin{thebibliography}{10}
\expandafter\ifx\csname url\endcsname\relax
  \def\url#1{\texttt{#1}}\fi
\expandafter\ifx\csname urlprefix\endcsname\relax\def\urlprefix{URL }\fi
\expandafter\ifx\csname href\endcsname\relax
  \def\href#1#2{#2} \def\path#1{#1}\fi

\bibitem{stauffer:book}
D.~Stauffer, A.~Aharony, Introduction to Percolation Theory, 2nd Edition,
  Taylor \& Francis, London, 1994.

\bibitem{grimmett:perc}
G.~Grimmett, Percolation, 2nd Edition, Springer, Berlin, 1999.

\bibitem{bollobas:06}
B.~Bollobas, O.~Riordan, Percolation, Cambridge University Press, Cambridge,
  2006.

\bibitem{nienhuis:domb}
B.~Nienhuis, Coulomb gas formulation of two-dimensional phase transitions, in:
  C.~Domb, J.~L. Lebowitz (Eds.), Phase Transitions and Critical Phenomena,
  Vol.~11, Academic Press, London, 1987, p.~1.

\bibitem{cardy:domb}
J.~L. Cardy, Conformal invariance, in: C.~Domb, J.~L. Lebowitz (Eds.), Phase
  Transitions and Critical Phenomena, Vol.~11, Academic Press, London, 1987,
  p.~55.

\bibitem{schramm:00}
O.~Schramm, Scaling limits of loop-erased random walks and uniform spanning
  trees, Israel J. Math. 118 (2000) 221.

\bibitem{smirnov:01}
S.~Smirnov, W.~Werner, Critical exponents for two-dimensional percolation,
  Math. Res. Lett. 8~(6) (2001) 729--744.
\newblock \href {http://dx.doi.org/10.4310/MRL.2001.v8.n6.a4}
  {\path{doi:10.4310/MRL.2001.v8.n6.a4}}.

\bibitem{smirnov:01a}
S.~Smirnov, Critical percolation in the plane: {C}onformal invariance,
  {Cardy}'s formula, scaling limits, C. R. Acad. Sci. Paris S\'{e}r. I Math.
  333~(3) (2001) 239--244.

\bibitem{duminil:12}
H.~Duminil-Copin, S.~Smirnov, Conformal invariance of lattice models, in:
  D.~Ellwood, C.~Newman, V.~Sidoravicius, W.~Werner (Eds.), Probability and
  Statistical Physics in Two and More Dimensions, Vol.~15 of Clay Mathematics
  Proceedings, American Mathematical Society, Providence, RI, 2012, pp.
  213--276.

\bibitem{fortuin:72a}
C.~M. Fortuin, P.~W. Kasteleyn, On the random-cluster model i. introduction and
  relation to other models, Physica 57~(4) (1972) 536--564.
\newblock \href {http://dx.doi.org/10.1016/0031-8914(72)90045-6}
  {\path{doi:10.1016/0031-8914(72)90045-6}}.

\bibitem{grimmett:book}
G.~Grimmett, The random-cluster model, Springer, Berlin, 2006.

\bibitem{wu:82a}
F.~Y. Wu, The {Potts} model, Rev. Mod. Phys. 54 (1982) 235--268.
\newblock \href {http://dx.doi.org/10.1103/RevModPhys.54.235}
  {\path{doi:10.1103/RevModPhys.54.235}}.

\bibitem{beffara:12}
V.~Beffara, H.~Duminil-Copin, The self-dual point of the two-dimensional
  random-cluster model is critical for \(q \ge 1\), Probab. Theory Rel.
  153~(3-4) (2012) 511--542.
\newblock \href {http://dx.doi.org/10.1007/s00440-011-0353-8}
  {\path{doi:10.1007/s00440-011-0353-8}}.

\bibitem{baxter:book}
R.~J. Baxter, Exactly Solved Models in Statistical Mechanics, Academic Press,
  London, 1982.

\bibitem{stanley:77}
H.~E. Stanley, Cluster shapes at the percolation threshold: and effective
  cluster dimensionality and its connection with critical-point exponents, J.
  Phys. A 10~(11) (1977) L211.
\newblock \href {http://dx.doi.org/10.1088/0305-4470/10/11/008}
  {\path{doi:10.1088/0305-4470/10/11/008}}.

\bibitem{coniglio:89}
A.~Coniglio, Fractal structure of ising and potts clusters - exact results,
  Phys. Rev. Lett. 62 (1989) 3054.
\newblock \href {http://dx.doi.org/10.1103/PhysRevLett.62.3054}
  {\path{doi:10.1103/PhysRevLett.62.3054}}.

\bibitem{deng:04}
Y.~J. Deng, H.~W.~J. Bl\"{o}te, B.~Nienhuis, Backbone exponents of the
  two-dimensional \(q\)-state {P}otts model: A {M}onte {C}arlo investigation,
  Phys. Rev. E 69 (2004) 026114.

\bibitem{xu:14}
X.~Xu, J.~Wang, Z.~Zhou, T.~M. Garoni, Y.~Deng, Geometric structure of
  percolation clusters, Phys. Rev. E 89~(1) (2014) 012120.

\bibitem{araujo:11a}
N.~A.~M. Araujo, K.~J. Schrenk, J.~S. Andrade, H.~J. Herrmann, Bridge
  percolation, Preprint arXiv:1103.3256\href {http://arxiv.org/abs/1103.3256}
  {\path{arXiv:1103.3256}}.

\bibitem{elci:15}
E.~M. El\c{c}i, M.~Weigel, N.~G. Fytas, Fragmentation of fractal random
  structures, Phys. Rev. Lett. 114~(11) (2015) 115701.

\bibitem{swendsen-wang:87a}
R.~H. Swendsen, J.~S. Wang, Nonuniversal critical dynamics in {Monte Carlo}
  simulations, Phys. Rev. Lett. 58 (1987) 86--88.
\newblock \href {http://dx.doi.org/10.1103/PhysRevLett.58.86}
  {\path{doi:10.1103/PhysRevLett.58.86}}.

\bibitem{chayes:98a}
L.~Chayes, J.~Machta, Graphical representations and cluster algorithms {II},
  Physica A 254 (1998) 477.

\bibitem{elci:13}
E.~M. El\c{c}i, M.~Weigel, Efficient simulation of the random-cluster model,
  Phys. Rev. E 88 (2013) 033303.
\newblock \href {http://dx.doi.org/10.1103/PhysRevE.88.033303}
  {\path{doi:10.1103/PhysRevE.88.033303}}.

\bibitem{sweeny:83}
M.~Sweeny, {M}onte {C}arlo study of weigted percolation clusters relevant to
  the potts model, Phys. Rev. B 27 (1983) 4445.

\bibitem{edwards:88a}
R.~G. Edwards, A.~D. Sokal, Generalization of the
  {Fortuin-Kasteleyn}-{Swendsen-Wang} representation and monte carlo algorithm,
  Phys. Rev. D 38 (1988) 2009.
\newblock \href {http://dx.doi.org/10.1103/PhysRevD.38.2009}
  {\path{doi:10.1103/PhysRevD.38.2009}}.

\bibitem{salas:97}
J.~Salas, A.~D. Sokal, Dynamic critical behavior of the {Swendsen-Wang}
  algorithm: The two-dimensional three-state {Potts} model revisited, J. Stat.
  Phys. 87~(1-2) (1997) 1--36.
\newblock \href {http://dx.doi.org/10.1007/BF02181478}
  {\path{doi:10.1007/BF02181478}}.

\bibitem{hu:14}
H.~Hu, H.~W.~J. Bl\"{o}te, R.~M. Ziff, Y.~Deng, Short-range correlations in
  percolation at criticality, Phys. Rev. E 90~(4) (2014) 042106.

\bibitem{grimmett:graphs}
G.~Grimmett, Probability on Graphs -- Stochastic Processes on Graphs and
  Lattices, Cambridge University Press, Cambridge, 2010.

\bibitem{steif:11}
J.~E. Steif, A mini course on percolation theory, Jyv\"{a}skyl\"{a} Lectures in
  Mathematics 3 (2011) 1.

\bibitem{essam:87}
J.~W. Essam, Connectedness and connectivity in percolation theory, in:
  A.~Barlotti, M.~Biliotti, A.~Cossu, G.~Korchmaros, G.~Tallini (Eds.), Annals
  of Discrete Mathematics (33) Proceedings of the International Conference on
  Finite Geometries and Combinatorial Structures, Vol. 144 of North-Holland
  Mathematics Studies, North-Holland, 1987, pp. 41--57.
\newblock \href {http://dx.doi.org/10.1016/S0304-0208(08)73047-0}
  {\path{doi:10.1016/S0304-0208(08)73047-0}}.

\bibitem{caselle:01}
M.~Caselle, F.~Gliozzi, S.~Necco, Thermal operators and cluster topology in the
  q -state potts model, J. Phys. A 34~(3) (2001) 351.

\bibitem{grimmett:95}
G.~Grimmett, The stochastic {Random-Cluster} process and the uniqueness of
  {Random-Cluster} measures, Ann. Probab. 23~(4).

\bibitem{hu:99}
C.-K. Hu, J.-A. Chen, N.~S. Izmailian, P.~Kleban, Geometry, thermodynamics, and
  finite-size corrections in the critical {P}otts model, Phys. Rev. E 60~(6)
  (1999) 6491.

\bibitem{ferdinand:69a}
A.~E. Ferdinand, M.~E. Fisher, Bounded and inhomogeneous {Ising} models. {I.}
  {Specific} heat anomaly of a finite lattice, Phys. Rev. 185 (1969) 832.
\newblock \href {http://dx.doi.org/10.1103/PhysRev.185.832}
  {\path{doi:10.1103/PhysRev.185.832}}.

\bibitem{salas:97a}
J.~Salas, A.~D. Sokal, Logarithmic corrections and finite-size scaling in the
  two-dimensional 4-state potts model, J. Stat. Phys. 88 (1997) 567.

\bibitem{nauenberg:80}
M.~{Nauenberg}, D.~J. {Scalapino}, {Singularities and Scaling Functions at the
  Potts-Model Multicritical Point}, Phys. Rev. Lett. 44 (1980) 837--840.
\newblock \href {http://dx.doi.org/10.1103/PhysRevLett.44.837}
  {\path{doi:10.1103/PhysRevLett.44.837}}.

\bibitem{cardy:80}
J.~L. {Cardy}, M.~{Nauenberg}, D.~J. {Scalapino}, {Scaling theory of the
  Potts-model multicritical point}, Phys. Rev. B 22 (1980) 2560--2568.
\newblock \href {http://dx.doi.org/10.1103/PhysRevB.22.2560}
  {\path{doi:10.1103/PhysRevB.22.2560}}.

\bibitem{deng:06}
Y.~Deng, X.~Yang, Finite-size scaling of energylike quantities in percolation,
  Phys. Rev. E 73~(6) (2006) 066116.

\bibitem{schmidt:13}
J.~M. Schmidt, A simple test on 2-vertex-and 2-edge-connectivity, Inform.
  Process. Lett. 113~(7) (2013) 241--244.

\bibitem{deng:14}
Y.~Deng, X.-W. Liu, J.~L. Jacobsen, Recursive percolation, arXiv preprint
  arXiv:1410.3603.

\bibitem{privman:privman}
V.~Privman, {Finite-Size} scaling theory, in: V.~Privman (Ed.), Finite Size
  Scaling and Numerical Simulation of Statistical Systems, World Scientific,
  Singapore, 1990, pp. 1--98.

\bibitem{salas:00}
J.~Salas, A.~D. Sokal, Universal amplitude ratios in the critical
  two-dimensional ising model on a torus, J. Stat. Phys. 98~(3-4) (2000)
  551--588, arXiv:9904038v1.

\bibitem{numrec}
W.~H. Press, S.~A. Teukolsky, W.~T. Vetterling, B.~P. Flannery, Numerical
  Recipes: The Art of Scientific Computing, 3rd Edition, Cambridge University
  Press, Cambridge, 2007.

\bibitem{coniglio:82}
A.~Coniglio, Cluster structure near the percolation transition, J. Phys. A 15
  (1982) 3829.

\bibitem{duplantier:89}
B.~Duplantier, H.~Saleur, Exact fractal dimension of of {2D} ising clusters,
  Phys. Rev. Lett. 63 (1989) 2536.

\bibitem{deng:10}
Y.~Deng, W.~Zhang, T.~M. Garoni, A.~D. Sokal, A.~Sportiello, Some geometric
  critical exponents for percolation and the random-cluster model, Phys. Rev. E
  81 (2010) 020102.
\newblock \href {http://dx.doi.org/10.1103/physreve.81.020102}
  {\path{doi:10.1103/physreve.81.020102}}.

\bibitem{beffara:11}
V.~Beffara, P.~Nolin, On monochromatic arm exponents for {2D} critical
  percolation, Ann. Probab. 39 (2011) 1286--1304.

\bibitem{vasseur:12}
R.~Vasseur, J.~L. Jacobsen, H.~Saleur, Logarithmic observables in critical
  percolation, J. Stat. Mech.: Theory Exp. 2012~(07) (2012) L07001.
\newblock \href {http://dx.doi.org/10.1088/1742-5468/2012/07/L07001}
  {\path{doi:10.1088/1742-5468/2012/07/L07001}}.

\bibitem{vasseur:14}
R.~Vasseur, J.~L. Jacobsen, Operator content of the critical {P}otts model in d dimensions
and logarithmic correlations, Nucl. Phys. B 880 (2014) 435.

\bibitem{grassberger:99a}
P.~Grassberger, Conductivity exponent and backbone dimension in 2-d
  percolation, Physica A 262~(3) (1999) 251--263.
\newblock \href {http://dx.doi.org/10.1016/S0378-4371(98)00435-X}
  {\path{doi:10.1016/S0378-4371(98)00435-X}}.

\bibitem{domany:81}
E.~Domany, D.~Mukamel, B.~Nienhuis, A.~Schwimmer, Duality relations and
  equivalences for models with {O(N}) and cubic symmetry, Nucl. Phys. B 190
  (1981) 279.

\bibitem{deng:07b}
Y.~Deng, T.~M. Garoni, W.~Guo, H.~W.~J. Bl\"{o}te, A.~D. Sokal, {Cluster
  Simulations of Loop Models on Two-Dimensional Lattices}, Phys. Rev. Lett.
  98~(12) (2007) 120601.
\newblock \href {http://dx.doi.org/10.1103/PhysRevLett.98.120601}
  {\path{doi:10.1103/PhysRevLett.98.120601}}.

\end{thebibliography}
